\documentclass[a4paper,twocolumn,11pt, accepted=2025-04-22]{quantumarticle}

\pdfoutput=1

\usepackage{xspace,amsmath,amsfonts,amsthm,amssymb,amsbsy}
\usepackage{graphicx}
\usepackage{amssymb}
\usepackage{color}
\usepackage{psfrag}
\usepackage{ifsym}
\usepackage{epstopdf}
\usepackage[usenames,dvipsnames]{xcolor}
\usepackage{bbold}
\usepackage{cancel}
\usepackage{braket}
\usepackage{rotating} %for sidewaystable
\usepackage{mathrsfs}
\usepackage{cite}

\usepackage{tikz-cd}
\usepackage{faktor}
\usepackage{orcidlink}

\usepackage{caption}
\usepackage{subcaption}
\usepackage[normalem]{ulem}
\usepackage{physics}
\usepackage{dsfont}
\usepackage{hyperref}
\hypersetup{
    colorlinks=true,
    linkcolor=blue,
    filecolor=red,      
    urlcolor=teal,citecolor=purple
}

\definecolor{orchid}{rgb}{0.85, 0.44, 0.84}
\definecolor{pink}{rgb}{1,0.078,0.57}
\definecolor{green}{rgb}{0,0.7,0}

\newcommand{\dg}{^{\dagger}}

\renewcommand{\H}{\mathcal{H}}
\newcommand{\B}{\mathcal{B}}

\newcommand{\Q}{\mathcal{Q}}

\newcommand{\T}{\mathsf{T}}

\newcommand{\V}{\mathcal{V}}
\newcommand{\W}{\mathcal{W}}

\newcommand{\F}{\mathcal{F}}

\newcommand{\E}{\mathcal{E}}

\newcommand{\M}{\mathcal{M}}

\newcommand{\CC}{\mathbb{C}}

\renewcommand{\d}{\mathrm{d}}
\newcommand{\id}{\mathds{1}}
\newcommand{\dt}{\mathrm{d}t}

\newcommand{\dist}{\operatorname{dist}}

\newcommand{\linspan}{\text{sp}}

\newcommand{\rk}[1]{\mathrm{rk}\left( #1 \right)}
\newcommand{\inner}[2]{(#1, #2)}

\begin{document}
\title{A geometrical description of non-Hermitian dynamics:
speed limits in finite rank density operators}

\author{Niklas H\"ornedal\,\orcidlink{0000-0002-2005-8694}\,}
\affiliation{Department of Physics and Materials Science, University of Luxembourg, L-1511 Luxembourg, G. D. Luxembourg}
\author{Oskar A. Pro\'sniak\,\orcidlink{0000-0003-4550-7561}\,}
\affiliation{Department of Physics and Materials Science, University of Luxembourg, L-1511 Luxembourg, G. D. Luxembourg}
\author{Adolfo del Campo\,\orcidlink{0000-0003-2219-2851}\,}
\affiliation{Department of Physics and Materials Science, University of Luxembourg, L-1511 Luxembourg, G. D. Luxembourg}
\affiliation{Donostia International Physics Center, E-20018 San Sebasti\'an, Spain}
\author{Aur\'elia Chenu\,\orcidlink{0000-0002-4461-8289}\,}
\affiliation{Department of Physics and Materials Science, University of Luxembourg, L-1511 Luxembourg, G. D. Luxembourg}

%\date{\today}

\maketitle
%\keywords{Differential geometry, non-Hermitian dynamics, shortcuts in open systems, speed limits}
\begin{abstract}
 Non-Hermitian dynamics in quantum systems preserves the rank of the state density operator. Using this insight, we develop a geometric framework to describe its time evolution. In particular, we identify mutually orthogonal coherent and incoherent directions and provide their physical interpretation. This understanding enables us to optimize the success rate of non-Hermitian driving along prescribed trajectories, with direct relevance to shortcuts to adiabaticity. Next, we explore the geometric interpretation of a speed limit for non-Hermitian Hamiltonians and analyze its tightness. We derive the explicit expression that saturates this bound and illustrate our results with a minimal example of a dissipative qubit.   

%\sout{Non-Hermitian dynamics in quantum systems preserves the rank of the state density operator. We use this insight to develop its geometrical description. In particular, we identify mutually orthogonal coherent and incoherent directions and give their physical interpretation. This understanding allows us to optimize the success rate for implementing non-Hermitian driving along prescribed trajectories. We show its significance for shortcuts to adiabaticity. We introduce the geometrical interpretation of a speed limit for non-Hermitian Hamiltonians and analyze its tightness. We derive the explicit expression that saturates such a speed limit and illustrate our results on a minimal example of a dissipative qubit.}

\end{abstract}
 
\section{Introduction}

Non-Hermitian dynamics naturally arises in a variety of applications in physics, engineering, and computer science \cite{Muga2004,Ashida2020}.
In quantum systems, it describes the evolution of the state of the system conditioned to remain in a given subspace and can be rigorously justified using the Feshbach projection approach.
Non-Hermitian evolution also arises in the context of continuous quantum measurements, in the description of a subensemble of quantum trajectories via post-selection, e.g., in the absence of quantum jumps (null-measurement conditioning) \cite{Carmichael2007vol2,jacobs_quantum_2014}. 
Non-Hermitian descriptions have also been used since the early days of quantum mechanics to describe decay rates phenomenologically. They find further applications in numerical methods for quantum dynamics and the theory of chemical reaction rates, among other examples \cite{Muga2004}.

In addition, an arbitrary differentiable trajectory of fixed-rank density matrices can be associated with a generator of evolution that is non-Hermitian. The associated equation of motion involves a nonlinear term to render the evolution trace-preserving \cite{alipour_shortcuts_2020}.

This work focuses on a geometric approach to non-Hermitian quantum evolution. The description of dynamics in geometric terms has a long history that has shed new light on the foundations of physics. It is central in information geometry, whether classical \cite{amari_information_2016} or quantum \cite{amari_methods_2007}. In quantum physics, it plays a key role in the understanding of quantum adiabatic dynamics and geometric phases \cite{Berry1984,Wilczek1989,Chruscinski2004}, proofs of the adiabatic theorem, and shortcuts to adiabaticity (STA) \cite{Berry2009, Torrontegui2013,guery-odelin_shortcuts_2019}. A geometric understanding of quantum dynamics is also instrumental in studying time-energy uncertainty relations and their refinement in the form of quantum speed limits, i.e., bounds on the minimum time for a process to unfold \cite{Deffner2017,Gong2022}.

Here, we describe how the set of reachable states under non-Hermitian dynamics forms a differentiable manifold. This observation allows for a geometrical interpretation of non-Hermitian dynamics. Specifically, the local effects of the dynamics can be described through a tangent space, which can be decomposed into a unitary and commutative subspaces. We show that these subspaces are maximally distinguishable with respect to any monotone metric.  This geometrical picture provides new insight into STAs generalizing counterdiabatic driving \cite{Demirplak03,Demirplak05,Demirplak08,Berry09} to open quantum systems \cite{Vacanti2014,alipour_shortcuts_2020,alipour_entropy-based_2022}. We then focus on the Bures metric and explicitly describe the geodesics on fixed-rank manifolds. We show applications to derive quantum speed limits for non-Hermitian systems and illustrate the control dynamics of a dissipative qubit. The notions of differential geometry that we use are described in Appendix~\ref{Appendix: geometry}.

\section{Non-Hermitian dynamics and fixed-rank manifold}
\label{Non-Hermitian dynamics and fixed-rank manifold}

In this section, we describe how non-Hermitian state dynamics arises as a special case of a continuous implementation of efficient measurements. We review how this dynamics gives rise to a group action on the state space by the general linear group and describe the differential structure of its orbits.

\subsection{Measurement and non-Hermitian dynamics}\label{sec: measurement}

From an axiomatic viewpoint, the possible transformations of closed quantum systems are commonly divided into two types: the deterministic unitary evolution, described by the Schrödinger equation, and the non-deterministic measurement process, described by projective measurement operators. More generally, one may choose to focus on an open system's effective dynamics induced by the unitary evolution of a system coupled to an ancilla and measurements of the latter. Such an effective dynamics is given, according to Stinespring's and Naimark's dilation theorems \cite{stinespring_positive_1955}, by a completely positive trace decreasing (CPTD) map that admits the Choi-Kraus representation \cite{Hayashi2016}
\begin{equation}\label{eq:1}
    \rho \mapsto \tilde{\Phi}(\rho):=\sum_{m=1}^N G_m \rho G_m^\dagger,
    \end{equation}
where $\rho$ is a reduced density matrix of the subsystem of interest and $G_m$ are the so-called Kraus operators fulfilling $\sum_{m=1}^N G_m^\dagger G_m\leq \mathds{1}$ \cite{nielsen_quantum_2010}.

For a composite system initialized in a product state with a pure ancilla state,  $\rho_0\otimes\ketbra{\phi_\textrm{A}}$,  the above CPTD map reduces to the action of a single Kraus operator. This protocol is known as an \emph{efficient measurement} 
\cite{wiseman_quantum_2009}. Equation~\eqref{eq:1} then gives a state conditioned on the outcome of that selective measurement, which is to be contrasted with trace-preserving nonselective measurements \cite{audretsch_mixed_2007}.

One can then pick another ancillary system in its pure state and iterate the procedure. In the simplest case,  the old ancilla is being reused since the projective measurement collapses its density matrix to a pure state. 
The limit in which the time of intermediate unitary evolution goes to zero is known as \emph{continuous measurement}.
The time evolution of the conditioned state can then be described by a map of the form $\rho \mapsto \tilde{\Phi}_t(\rho):= G(t)\rho G^\dagger(t)$. Having chosen an initial state $\rho_0$, we will talk about its trajectory $\rho(t):=\tilde{\Phi}_t(\rho_0)$.
Given that $G(t)$ is differentiable and invertible, the dynamical map $\tilde{\Phi}_t$ can be associated with the following dynamics
\begin{equation}
    \dot{\rho}(t) = -i\left(K(t)\rho(t) - \rho(t) K^\dagger(t)\right),
\end{equation}
where $K(t) = i\dot{G}(t)G^{-1}(t)$ acts as a generator of the evolution operator $G(t)$ and need not be Hermitian. Physically, we can think of the map $\tilde{\Phi}_t$ as a channel describing a continuous information flow from the system to the observer \cite{fuchs_information-tradeoff_2001, nielsen_characterizing_2001}. 

\subsection{States with fixed rank: manifold \texorpdfstring{$\Q_r$}{\265}}
The invertibility condition above implies that any state has a non-zero probability of passing through the channel $\tilde{\Phi}_t$. But since the map $\tilde{\Phi}_t$ is trace-decreasing, the state passes through the channel only with probability ${\rm Tr}[\tilde{\Phi}_t(\rho)]$.
Consequently, the output of the map $\tilde{\Phi}_t$ does not need to be a physical density matrix. The remedy to this problem is to normalize the output.

Formally, let us denote by $\Q$ the set of all density operators acting on a $n$-dimensional Hilbert space $\H_n$. The non-Hermitian evolution leading to a normalized state, described above, takes the general form
\begin{equation}
\label{eq: normalised map}
    \Phi_t: \Q \to \Q, \quad \Phi_t(\rho):=\frac{G(t)\rho G^\dagger (t)}{\Tr(G(t)\rho G^\dagger (t))},
\end{equation}
where, as before, $G(t)$ is a smooth function of $t$ valued in invertible operators, i.e., elements of the general linear group $\mathrm{GL}(n,\CC)$. Note that $G(t)$ can be unitary, but we do not impose it. Since $G(t)$ is necessarily of full rank, when acting on density matrices $\rho$, it preserves their rank,\footnote{Notice that \eqref{eq: normalised map} is also the most general form of a normalized CPTD map that preserves the rank of any positive operator.} i.e., if $\rk{\rho}=r$ then $\rk{\Phi_t(\rho)}=r$.\footnote{Although the rank of the evolving state does not change, its  trace distance to states with a different rank can decrease rapidly, but never reach zero in a finite time.} Hence, to study the evolution of a density operator initially of rank $r$, we may restrict our attention to the subset $\Q_r\subset \Q$ of density operators with the same rank. 

\begin{figure}
     \centering
     \includegraphics[width=1.0\columnwidth]{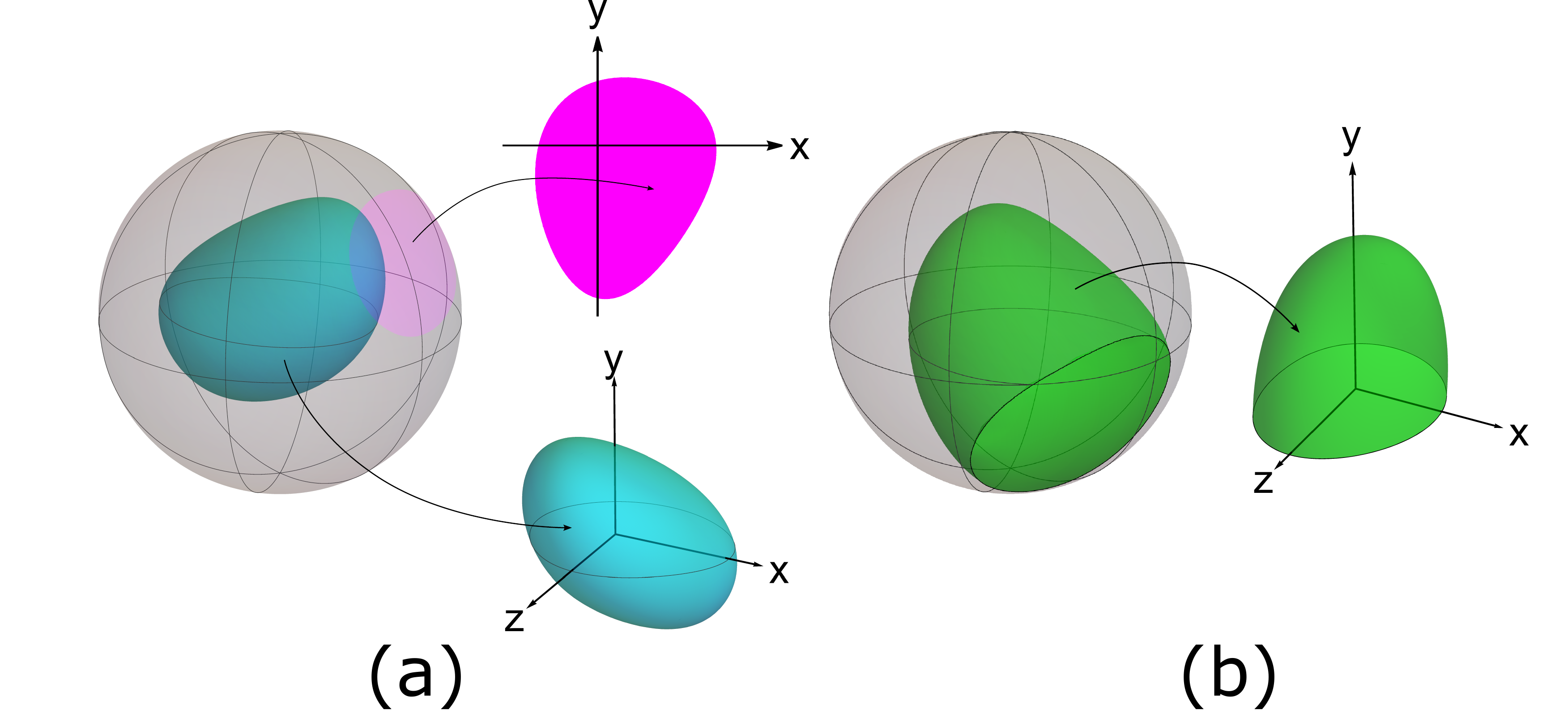}
          \vspace{1ex}
     \caption{(a) The surface and the interior of the Bloch's ball respectively form $2-$ and $3-$dimensional manifolds. (b) The Bloch's ball itself does not form a standard manifold as open sets touching its surface cannot be mapped to open sets in $\mathbb{R}^d$ for any $d$.\protect\footnotemark[3]}
     \label{fig: manifold}
 \end{figure}

This observation allows us to introduce a differential structure. Indeed, the set $\Q_r$ is naturally a $(2nr -r^2-1)$-dimensional 
manifold with a differential structure inherited from the space of $n \times n$ complex matrices, isomorphic to $\mathbb{R}^{2n^2}$, in which it can be smoothly embedded \cite{grabowski_geometry_2005}. This is not the case for $\Q$, which is not a manifold in the usual sense (see Fig.~\ref{fig: manifold}). However, it can also be endowed with a generalized differential structure, which we comment on in the conclusion.

\footnotetext{The Bloch's ball forms a manifold with boundaries. Defining a differential structure on such manifolds does not pose difficulties. However, higher-dimensional systems can no longer be described as such, and extending the presented results to them is not straightforward.}

\subsection{Velocities on \texorpdfstring{$\Q_r$}{\265}: tangent space \texorpdfstring{$\mathsf{T}_\rho\Q_r$}{\265}}

\begin{figure}
    \centering
    \includegraphics[width=0.9\columnwidth]{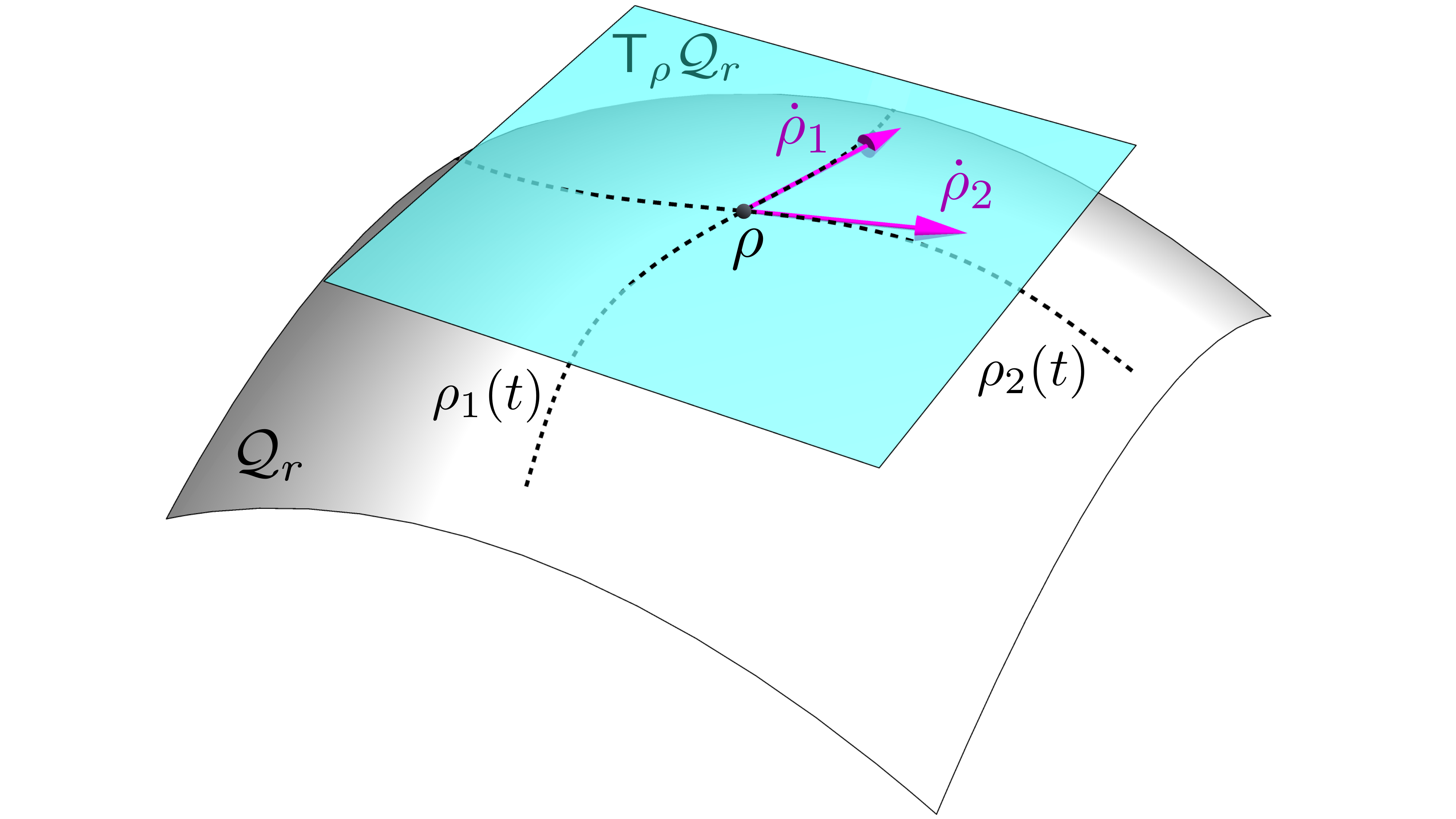}
    \caption{The manifold $\Q_r$ (white/grey surface). We illustrate two evolution curves on this manifold, $\rho_1(t)$ and $\rho_2(t)$, that pass through $\rho$; the arrows in magenta show the corresponding tangent vectors at this point, $\dot{\rho}_1$ and $\dot{\rho}_2$. These tangent vectors are elements of the tangent space at $\rho$, denoted by $\mathsf{T}_\rho\Q_r$, and depicted as the cyan plane.}
 \label{fig:tangent}
\end{figure}

Consider any curve describing the evolution of a quantum state $\rho(t)$, governed by the map \eqref{eq: normalised map} and passing through the state $\rho$ at some time that we take to be zero. The tangent to this curve at $\rho$ can be associated with its time derivative $\dot{\rho}\equiv \dot{\rho}(0)$. Importantly, the set of all such tangents forms the \textit{tangent space} $\mathsf{T}_\rho \Q_r$, illustrated in Fig.~\ref{fig:tangent}.

The tangent vector can be expressed in different ways. 
First, considering that any two states in $\Q_r$ can be connected by the group action
\begin{equation}
    \begin{split}
    &\Phi:\Q_r\times\mathrm{GL}(n,\CC)\to\Q_r,\\
    &\rho \mapsto \Phi(\rho,G):=\frac{G\rho G^\dagger}{\Tr( G \rho G^\dagger ) }.
    \end{split}
\end{equation}
Any tangent vector at $\rho$ ($\dot{\rho} \in \mathsf{T}_\rho\Q_r$) follows as
\begin{equation}\label{eq: rho dot}
    \dot{\rho} = -i (K \rho - \rho K^\dagger ) + i \Tr((K-K^\dagger )\rho)\rho, 
\end{equation}
where $K\in\mathfrak{gl}(n,\CC)$ is an element of the Lie algebra associated with $\mathrm{GL}(n,\CC)$, consisting of all $n\times n$ matrices \cite{ciaglia_quantum_2020}.

Alternatively, one can decompose the generator of the dynamics into its Hermitian and anti-Hermitian parts, namely $K=H-i\Gamma$ with $H = (K+K^\dagger)/2$ and $\Gamma = (K^\dagger-K)/(2i)$. The tangent vector then takes the form \cite{Brody2012}
\begin{equation}
\label{eq: rho dot H and Gamma}
    \dot{\rho} = -i \left[ H,\rho \right] - \lbrace \Gamma,\rho \rbrace +2\Tr(\Gamma\rho)\rho.
\end{equation}
This equation arises in continuous quantum measurements in the absence of quantum jumps, i.e., under null-measurement conditioning \cite{Carmichael2007vol2}. It also occurs in spectral filtering, which modifies the dynamics and enhances the dynamical manifestations of quantum chaos \cite{cornelius_spectral_2022,MatsoukasRoubeas2023,MatsoukasBeau2023}.

The non-linear term appearing in \eqref{eq: rho dot} and \eqref{eq: rho dot H and Gamma} is due to the renormalization of the state to preserve its norm. Note that a master equation similar to \eqref{eq: rho dot H and Gamma}  has also been considered to describe classical dissipative systems \cite{das_density_2022}.
We discuss this ``dynamical decomposition" further in Section~\ref{gradient}, but first focus on a different ``static decomposition" of the tangent vectors, which refers to the geometry of the tangent space itself.

\section{Geometry of non-Hermitian trajectories}\label{sec:UandC}

\subsection{Coherent and incoherent directions: spaces \texorpdfstring{$\mathsf{U}_\rho\Q_r$}{\265} and \texorpdfstring{$\mathsf{C}_\rho\Q_r$}{\265} }

\begin{figure}
     \centering
     \includegraphics[trim=140 0 140 0,clip,width=0.8\columnwidth]{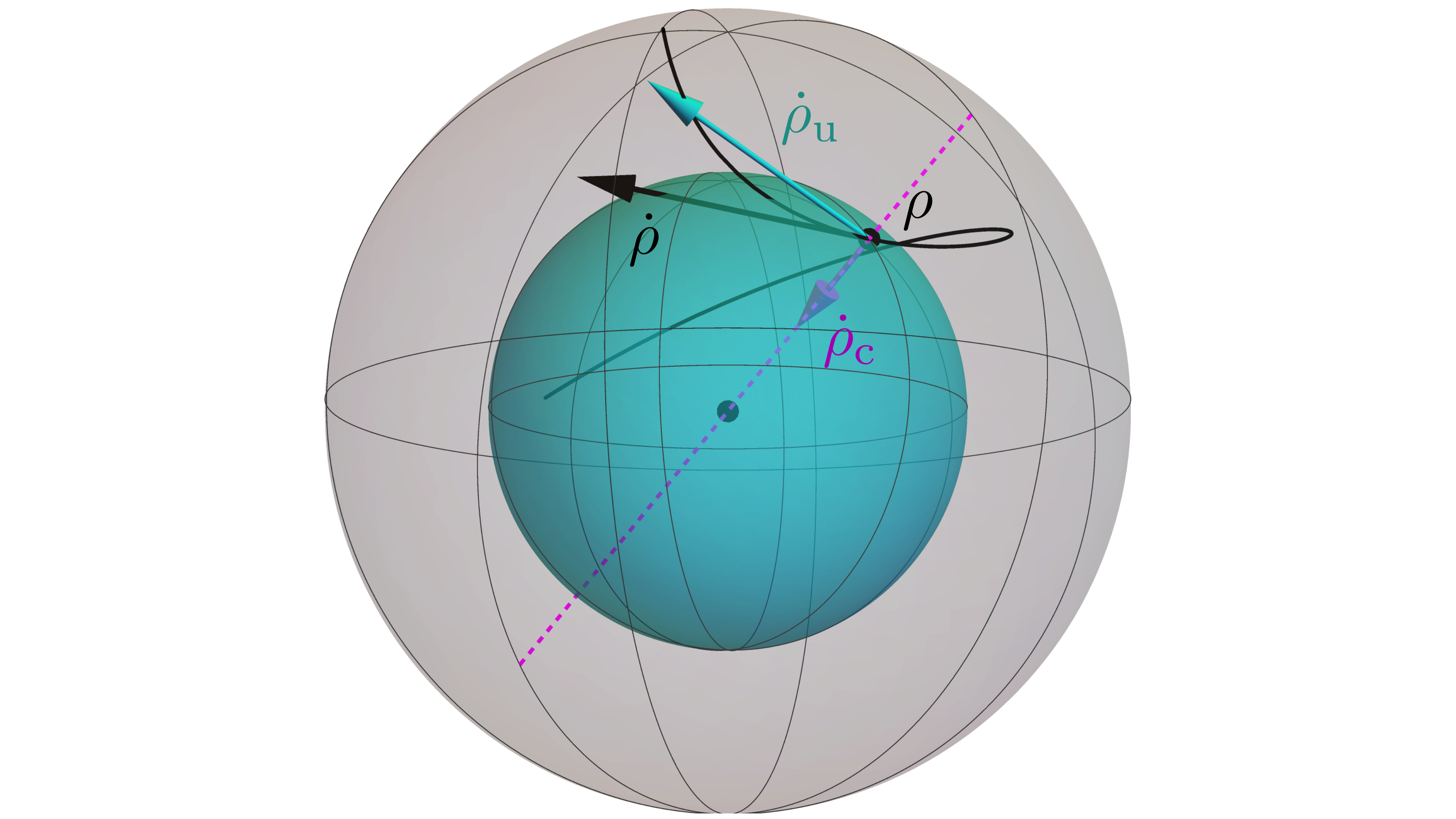}
     \caption{At every point of the trajectory in the interior of the Bloch's ball, the tangent to it (black arrow) can be decomposed into commutative (magenta) and unitary (cyan) components.}
     \label{fig: C and U}
 \end{figure}

Using the spectral decomposition of the density matrix $\rho = \sum_{k} \lambda_k \Pi_k$, where $\Pi_k$ are spectral projections and $\lambda_k$ are distinct eigenvalues, any vector tangent to $\Q_r$ at this point can be decomposed into $\dot{\rho} = \dot{\rho}_\textrm{u} + \dot{\rho}_\textrm{c}$, with 
\begin{subequations}
    \begin{align}
        \label{coherent decomp 1}
   &\dot{\rho}_\textrm{u} =  \sum_{k\neq m} \Pi_k\dot{\rho}\Pi_m,\\
   &\dot{\rho}_\textrm{c} = \sum_k \Pi_k\dot{\rho}\Pi_k.
    \end{align}
\end{subequations}
This decomposition, illustrated in Fig.\ \ref{fig: C and U}, has been explored before in the context of information geometry \cite{hubner_computation_1993, hasegawa_divergence_1993, petz_riemannian_1996, dittmann_connections_1999, andersson_holonomy_2019}. It has also been employed to bound the speed of evolution \cite{girolami_how_2019, funo_speed_2019, garcia-pintos_unifying_2022} and to identify work and heat in quantum thermodynamics \cite{Alipour2022}. Moreover, this framework extends naturally to arbitrary observables \cite{garcia-pintos_unifying_2022}. 
%\sout{This decomposition, illustrated in Fig. \ref{fig: C and U}, has been used in the context of information geometry \cite{hubner_computation_1993, hasegawa_divergence_1993, petz_riemannian_1996, dittmann_connections_1999, andersson_holonomy_2019}. It was used in Refs. \cite{girolami_how_2019,funo_speed_2019,garcia-pintos_unifying_2022} for bounding the speed of the evolution, and in \cite{Alipour2022} for the identification of work and heat in quantum thermodynamics. It can be generalized to arbitrary observables \cite{garcia-pintos_unifying_2022}.}

It follows that the tangent space $\mathsf{T}_\rho \Q_r$ can be decomposed into a direct sum of unitary and commutative subspaces \cite{amari_methods_2007,bengtsson_geometry_2006, andersson_holonomy_2019}
respectively defined as
\begin{subequations}
\begin{align}
 \!\!  \mathsf{U}_\rho\Q_r &= \lbrace \dot{\rho}\in \mathsf{T}_\rho\Q_r | \dot{\rho}= -i[H,\rho], H=H^\dag{} \rbrace,\\
  \!\! \mathsf{C}_\rho\Q_r &= \lbrace \dot{\rho}\in \mathsf{T}_\rho\Q_r | [\dot{\rho},\rho]=0 \rbrace.
\end{align}
\end{subequations}
It is easy to check that $\dot{\rho}_\textrm{u} \in \mathsf{U}_\rho\Q_r$ and $\dot{\rho}_\textrm{c} \in \mathsf{C}_\rho\Q_r$, and that these spaces are linearly independent. 

Remember that elements of $\mathsf{T}_\rho \Q_r$ can be thought of as tangents at $\rho$ to all the smooth curves in $\Q_r$ passing through this point at time zero. The set of allowed curves can be restricted to those that form smooth submanifolds, i.e., curves without cusps. It is then possible to define an eigen-decomposition along each such path $\rho(t)=\sum_{k=1}^{r} \lambda_k(t) \ketbra{\lambda_k(t)}$, where the one-dimensional projectors $\ketbra{\lambda_k(t)}$ depend smoothly on time.\footnote{Here the eigenvalues $\lambda_k$ are no longer assumed to be distinct. Higher dimensional projectors may be discontinuous at times when the degeneracy changes.} The unitary and commutative components of $\dot{\rho}=\dot{\rho}(0)$ take then a simple form
\begin{subequations}
\label{coherent decomp 2}
\begin{align}
   \dot{\rho}_\textrm{u} &=  \sum_{k} \lambda_k (\ketbra*{\dot{\lambda}_k}{\lambda_k}+\ketbra*{\lambda_k}{\dot{\lambda}_k}),\\
   \dot{\rho}_\textrm{c} &= \sum_k \dot{\lambda}_k \ketbra{\lambda_k}.
\end{align}
\end{subequations}

Physically, $\mathsf{U}_\rho \Q_r$ and $\mathsf{C}_\rho \Q_r$ can be thought of as ``coherent" and ``incoherent" directions, respectively, and we will henceforth refer to them as such \cite{garcia-pintos_unifying_2022,funo_speed_2019}. The evolution along the former does not change populations in the instantaneous eigenbasis of the state but generates coherence by contrast to the evolution along the latter.

Note that pure states ($r=1$) have a trivial commutative subspace, $\mathsf{C}_\rho\Q_{1} = \{0\}$, which agrees with the fact that the only quantum operations preserving purity are unitaries.

\subsection{Classical direction: subspace \texorpdfstring{$\mathsf{Class}_\rho\Q_r$}{\265} and complementary \texorpdfstring{$\mathsf{L}_\rho\Q_r$}{\265}}

The commutative subspace can be decomposed further when the spectrum of the state is degenerate \cite{andersson_holonomy_2019}---i.e.,\ $\Tr(\Pi_k)>1$ for at least one $k$. Indeed, the incoherent component $\dot{\rho}_\textrm{c}\in\mathsf{C}_\rho \Q_r$ can be expressed as $\dot{\rho}_\textrm{c} = \dot{\rho}_\textrm{cl} + \dot{\rho}_\textrm{l}$, with
\begin{subequations}
    \begin{align}
        \dot{\rho}_\textrm{cl} &= \sum_k \frac{\Tr(\Pi_k\dot{\rho})}{\Tr(\Pi_k)}\Pi_k,\\
        \dot{\rho}_\textrm{l} &= \sum_k \left( \Pi_k\dot{\rho}\Pi_k -\frac{\Tr(\Pi_k\dot{\rho})}{\Tr(\Pi_k)}\Pi_k\right).
    \end{align}
\end{subequations}
These directions form what we call the \textit{classical} $\mathsf{Class}_\rho\Q_r$ and the \textit{lifting} $\mathsf{L}_\rho\Q_r$ subspaces correspondingly. 

To sum up, the tangent vector can be decomposed as 
\begin{equation}\label{eq11}
    \dot{\rho}= \dot{\rho}_\textrm{u} + \dot{\rho}_\textrm{cl}+ \dot{\rho}_\textrm{l}.
\end{equation}
The lifting direction $\dot{\rho}_\textrm{l}$ is responsible for lifting the degeneracy of the state and is, for example, relevant in symmetry breaking in adiabatic processes. Evolution along the unitary and lifting directions can generally lead to a change of eigenbasis. As such, they can be viewed as non-classical. By contrast, the classical direction $\dot{\rho}_\textrm{cl}$ can be interpreted as classical information processing since we can, in a consistent way, map $\rho$ to a probability vector and consider the evolution along $\dot{\rho}_\textrm{cl}$ as arising through stochastic transformations. Note that ``classical" evolution assumes the eigenbasis is fixed and that only the evolution along the above-defined classical directions is guaranteed to leave the basis unchanged.
Lastly, it is worth noting that $\dot{\rho}_\textrm{u}$ preserves the von Neumann entropy of the state while $\dot{\rho}_\textrm{l}$ strictly decreases it.

\subsection{Maximal distinguishability of the coherent and incoherent directions}

Let us, for a moment, restrict our attention to the manifold of full-rank states $\Q_n$. It turns out that the coherent and incoherent directions are maximally distinguishable, or orthogonal, with respect to any monotone Riemannian metric on $\Q_n$. These are the metrics whose geodesic distance between any two density operators in $\Q_n$ either decreases or stays the same under completely positive trace preserving (CPTP) transformations. It can be argued that any meaningful notion of statistical distance on $\Q_n$ should have this property. Intuitively, two density matrices should not become more distinguishable after post-processing the data they describe.

Any monotone metric on the manifold $\Q_n$ is an inner product defined locally and associated with the following norm \cite{morozova_markov_1991, petz_monotone_1996}
\begin{equation}
\label{eq: monotone metric}
\begin{split}
    &\norm{\cdot}: \mathsf{T}_\rho \Q_n \to \mathbb{R}, \\
    &\norm{\dot{\rho}} = \sqrt{\sum_{j,k} |\dot{\rho}_{jk}|^2c(\lambda_j,\lambda_k)},
\end{split}
\end{equation}
where $\dot{\rho}_{jk} = \bra{\lambda_j}\!\dot{\rho}\!\ket{\lambda_k}$. The function $c(\lambda, \mu)$ is symmetric and homogeneous of order $-1$, i.e., $c(\alpha \lambda,\alpha \mu)=\alpha^{-1}c(\lambda,\mu)$. Its choice defines the metric. For example, choosing $c(\lambda, \mu)= \frac{2}{\lambda + \mu}$ leads to the Bures metric, widely used in quantum theory \cite{Hayashi2016}.

Using the definition of the monotone metric and the polarization identity, it follows that the coherent and incoherent directions are mutually orthogonal -- see Appendix \ref{app: orth} for a detailed proof. This maximal distinguishability can be visualized on the Bloch ball for a qubit, as seen in Fig.~\ref{fig: C and U}. Similarly, the lifting and classical subspaces can also be shown to be orthogonal % with respect to one another 
(Appendix \ref{app: orth}). These results reveal a hidden rigidity condition that all monotone metrics need to respect.

The orthogonality of coherent and incoherent directions with respect to all monotone metrics can be proven as above for $\Q_n$ only, as only on the manifold of full-rank states are the norms associated with these metrics known to be of the form $\eqref{eq: monotone metric}$. Moreover, the monotonicity of the metric is relevant for such states only, as information is indisputably gained under rank preserving dynamics \eqref{eq: normalised map} in open systems, due to efficient measurements described above.\footnote{The set of CPTP maps will in general preserve the rank only for full rank states.} However, the Bures metric can be extended to the manifold of states with rank $r<n$ \cite{dittmann_riemannian_1995,ciaglia_jordan_2020}. Specifically, on that manifold, the Bures metric $\inner{\cdot}{\cdot}_\textsc{b}: \mathsf{T}\Q_r\times\mathsf{T}\Q_r \to \mathbb{R}$ takes the form
\begin{equation}\label{eq:Bures}
    (\dot{\rho}, \dot{\rho}')_\textsc{b} = \frac{1}{2}\Tr(L\dot{\rho}'),
\end{equation}
where $L$ is known as a symmetric logarithmic derivative (SLD) and is any solution to the equation $L\rho +\rho L = \dot{\rho}$ \cite{Braunstein1994,dittmann_riemannian_1995}.\footnote{Note that we do not include any factor of two in this definition.} The coherent and incoherent directions are also orthogonal in $\Q_r$ with respect to this extended Bures metric.

\subsection{Hamiltonian and gradient vector fields}
\label{gradient}
Consider now a vector field representing a dynamics generated by a non-Hermitian operator $K = H-i\Gamma$. Equation \eqref{eq: rho dot H and Gamma} shows a decomposition of the vector field into two parts where the first term comes from the Hermitian part of $K$
\begin{equation}
    \dot{\rho}_H= -i[H,\rho],
\end{equation}
and the other two terms from the anti-Hermitian part
\begin{equation}
    \dot{\rho}_\Gamma = -\lbrace \tilde{\Gamma}_\rho,\rho\rbrace,
\end{equation}
where we have introduced the state-dependent operator $\tilde{\Gamma}_\rho = \Gamma - \Tr(\Gamma \rho)$.
The latter can be seen as a gradient of the function 
$\expval{\Gamma}:\Q_r\to\mathbb{R},\quad \expval{\Gamma}_\rho= \Tr(\Gamma \rho)$
with respect to the Bures metric \cite{ciaglia_quantum_2020}.

\begin{figure}
     \centering
     \includegraphics[width=1.0\columnwidth]{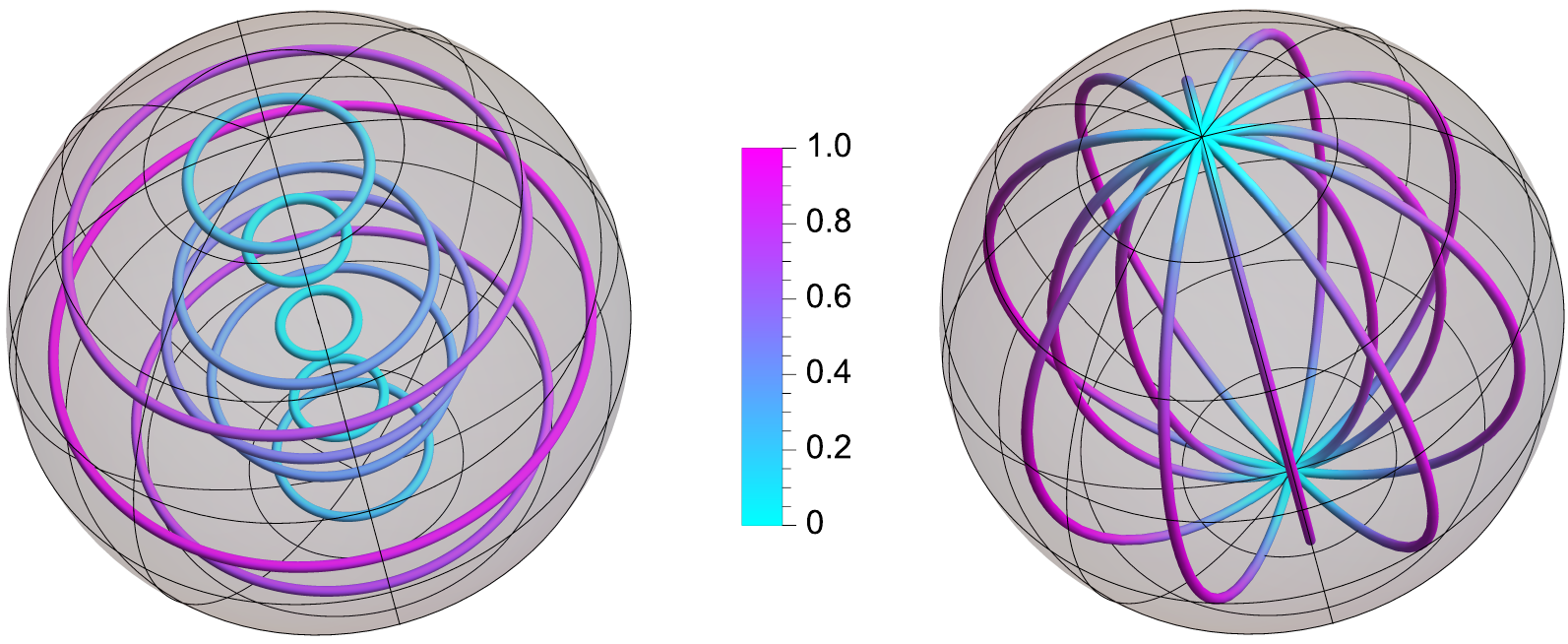}
     \caption{Vector fields $-i[A,\rho]$ (on the left) and $\{A,\rho\}$ (on the right) on the Bloch ball for a qubit, with $A = \sigma_z$. Colors represent the speed with respect to the Bures metric. The dynamics $-i[A,\rho]$ preserves the expectation values of $A$ as it rotates the sphere while $\{A,\rho\}$ flows from the south pole to the north pole along the direction of maximal increase.}
     \label{fig:gradient}
 \end{figure}

We recall that the gradient of any scalar field $f:\Q_r \to \mathbb{R}$ is a vector field $\nabla f:\Q_r\to\mathsf{T}\Q_r$ satisfying $\forall \rho\in \Q_r, \forall \dot{\rho}\in\mathsf{T}_\rho\Q_r: \,(\nabla f(\rho), \dot{\rho})_\textsc{b} = \d f(\dot{\rho})$,
where $\d f$ is the differential of $f$. In the case when $f$ is the expectation value of an observable $A$, $f(\rho)=\textrm{Tr}(A\rho)=\expval{A}_\rho$, then $\d \expval{A}_\rho=\Tr(A \dot{\rho})$. 
Interestingly, the fields $-i[A,\rho]$ and $\{A,\rho\}$ are everywhere orthogonal with respect to the Bures metric. The first one generates a rotation that preserves the expectation value of $A$, while the latter generates a flow that moves along the direction of maximal change and cuts each level hypersurface\footnote{A level hypersurface is a set on which the function takes a constant value.} orthogonally; see Fig.~\ref{fig:gradient}. For $A$ with a non-degenerate spectrum, the latter gives rise to $1$ stable, $1$ unstable, and $n-2$ hyperbolic fixed points corresponding to the observable's eigenstates. More generally, the vector fields $-i[H,\rho]$ and $\{\tilde{\Gamma}_\rho,\rho\}$ are everywhere orthogonal when $[H,\Gamma] = 0$.

Note that while the decomposition into coherent and incoherent directions purely refers to the differential structure of $\Q_r$, the decomposition into Hermitian and anti-Hermitian directions is dynamical, as it depends on the chosen generator of the dynamics. Moreover, while $\dot{\rho}_H\in\mathsf{U}_\rho\Q_r$, in general $\dot{\rho}_\Gamma \in \mathsf{U}_\rho\Q_r\oplus \mathsf{C}_\rho\Q_r$. 

\subsection{Maximal success rate}
\label{optimal}
When the non-Hermitian generator arises from the continuous measurements (see Sec.~\ref{sec: measurement}), the Hermitian operator $\Gamma$ is necessarily positive-semi definite due to the monotonic decrease of the state's norm. The norm decreases at a rate $\gamma=2\Tr(\Gamma\rho)$, which is the non-linear contribution in \eqref{eq: rho dot H and Gamma}.

Let us pick an initial state and evolve it according to \eqref{eq: rho dot H and Gamma}. At each time step, the anti-Hermitian part of the generator naturally splits into two parts $\Gamma = \Gamma_\textrm{u} + \Gamma_\textrm{c}$, where $\Gamma_\textrm{u} = \sum_{j\neq k}\Pi_j\Gamma\Pi_k$ induces motion in the coherent directions and $\Gamma_\textrm{c} = \sum_{k}\Pi_k\Gamma\Pi_k$ in the incoherent directions. Since $\Gamma_\textrm{u}$ generates the dynamics along coherent directions, by definition, there exists a (state-dependent) Hamiltonian generating the same dynamics. Therefore, by exchanging $\Gamma_\textrm{u}$ for this Hamiltonian, one can mitigate the decrease in norm along the given trajectory, as there is no probability decay for purely unitary dynamics. This leads to finding a maximal success rate, as we explain below.

Let us first observe that after correcting for the change in norm \eqref{eq: rho dot H and Gamma}, the dynamics remains unaffected by a shift
%\sout{is independent of the joint shifts of the eigenvalues} 
of the spectrum of $\Gamma$. However, the decay rate is not independent of these shifts and, under the constraint $\Gamma\geq0$, is minimized when the smallest eigenvalue of $\Gamma$ equals zero. We will view this as trivial optimization and compare rates for operators already shifted in this manner. 

Let $\mu_\textrm{c}^\textrm{(min)}$ and $\mu^\textrm{(min)}$ denote the smallest eigenvalues of $\Gamma_\textrm{c}$ and $\Gamma$ respectively. In light of the above comment, we set $\mu^\textrm{(min)} = 0$. Since $\Gamma_\textrm{c}$ and $\rho$ commute, they share a common eigenbasis. For any common eigenvector $\ket{\mu_\textrm{c}}$ of $\Gamma_\textrm{c}$ and $\rho$, we have $\bra{\mu_\textrm{c}}\!\Gamma\!\ket{\mu_\textrm{c}} = \bra{\mu_\textrm{c}}\!\Gamma_\textrm{c}\!\ket{\mu_\textrm{c}}$, which implies that $\mu_\textrm{c}^\textrm{(min)} \geq 0$.
The shifted operator $\Gamma_\textrm{opt} = \Gamma_\textrm{c} - \mu_\textrm{c}^\textrm{(min)} \id$ is positive ($\Gamma_\textrm{opt}\geq 0$) and satisfies
\begin{equation}
    \Tr(\Gamma_\textrm{opt}\: \rho) = \Tr(\Gamma\rho) - \mu_\textrm{c}^{\textrm{(min)}} \leq \Tr(\Gamma\rho).\label{eq: optimal Gamma}
\end{equation}
This inequality is saturated iff the kernel of $\Gamma_\textrm{c}$ is contained within the kernel of $\Gamma$\footnote{The kernel of a linear operator $A$ on a vector space $\V$ is the subspace $\textrm{ker}A\subset \V$ defined by $\textrm{ker}A = \{v\in\V\,\vert\, Av = 0\}$.}---something that never happens for full-rank qubit states when $\Gamma_\textrm{u} \neq 0$. 

We modify the Hermitian operator $H$ by adding the term $H' = - i\sum_{j\neq k}\frac{\lambda_j + \lambda_k}{\lambda_j - \lambda_k}\Pi_k\Gamma \Pi_j$,  which can be shown to satisfy $-i[H',\rho] = \{\Gamma_\textrm{u},\rho\}$. The modified Hermitian operator $H_\textrm{opt} = H + H'$ then generates the same evolution as $H-i\Gamma_\textrm{u}$. Consequently, $K_\textrm{opt} = H_\textrm{opt} -i\Gamma_\textrm{opt}$ generates the same evolution as $K$, but with a smaller decrease of the state's norm. Importantly, while the construction is set up to locally optimize the decay rate, it also maximizes the total success rate over the prescribed trajectory---see App.~\ref{app: opti} for a detailed argument.

To conclude, the non-Hermitian dynamics that maximizes the success rate for obtaining the desired state trajectory is achieved when the anti-Hermitian part of the generator commutes with the instantaneous state.
We stress that both $H_\textrm{opt}$ and $\Gamma_\textrm{opt}$ are constructed using the spectral projectors $\Pi_k$ of the instantaneous state and therefore inherently depend on the evolved state itself.

\subsection{Shortcuts to adiabaticity}
The geometrical interpretation presented in the preceding sections may be applied to STA in open systems. Under slow driving of a physical system, and assuming that the relevant spectral gaps remain open, adiabatic theorems guarantee that the instantaneous eigenstate populations remain constant \cite{Kato1950,Jansen2007,sarandy_adiabatic_2005,joye_adiabatic_2022}.
Shortcuts to adiabaticity are protocols that implement the evolution between a given initial state and an adiabatically reachable final state but in finite time, that is, without relying on slow driving. Originally developed for unitary evolutions, for example, by counterdiabatic driving \cite{Demirplak03,Berry09}, they have been extended to open systems \cite{Vacanti2014,alipour_shortcuts_2020}. Especially, STA for a given fixed-rank trajectory may be associated with an evolution that is, in general, generated by a non-Hermitian Hamiltonian.

Here, we show how to construct the STA generator of the trajectory
\begin{equation}
\label{sta dyn}
    \rho(t) = e^{-\beta H_0(t)}/Z(t),
\end{equation}
governed by the Hermitian time-dependent Hamiltonian $H_0(t)$, with inverse temperature $\beta>0$,  and $Z(t) = \textrm{Tr}(e^{-\beta H_0(t)})$ the partition function.  Note that any full-rank density matrix ($r=n$) can be written in this form. 
%\oap{\sout{Given the finite temperature, the rank of the state equals that of $H_0(t)$.}}
The velocity of \eqref{sta dyn} is given by $\dot{\rho}(t) = \dot{\rho}_\textrm{u}(t) + \dot{\rho}_\textrm{c}(t)$, where
\begin{subequations}
\begin{align}
\dot{\rho}_\textrm{u}(t) &= -i[H_{_{\textrm{CD}}}(t),\rho(t)], \\
\dot{\rho}_\textrm{c}(t) &= -\frac{\dot{Z}(t)}{Z(t)^2}\rho(t) -\frac{\beta}{Z(t)}\sum_{k=1}^n \dot{E}_{k}e^{-\beta E_{k}} \ketbra{E_{k}}
\end{align}
\end{subequations}
where we have used the eigendecomposition $(H_0(t) - E_{k})\ket{E_{k}} =0$. The operator $H_{_{\rm CD}}(t) = H_0(t) + H_1(t)$ is the counterdiabatic Hamiltonian with $H_1(t) = i\sum_{k=1}^n (\ketbra*{\dot{E}_{k}}{E_{k}} + \braket*{\dot{E}_{k}}{E_{k}}\ketbra{E_{k}})$. The subscripts of the two components were chosen deliberately as it is easy to check that $\dot{\rho}_\textrm{u}(t) \in \mathsf{U}_{\rho(t)}\Q_r$ and $\dot{\rho}_\textrm{c}(t) \in \mathsf{C}_{\rho(t)}\Q_r$. 

The same trajectory can be generated by the non-Hermitian generator $K(t) = H_{_\textrm{CD}}(t) - i\Gamma(t)$, where $\Gamma(t) = -\frac{\beta}{2}\sum_{k=1}^n \dot{E}_{k}\ketbra{E_{k}} - \frac{1}{2}\dv{t}\log{Z}(t)$. Since $[\Gamma(t),\rho(t)] = 0$, this implementation optimizes the success rate, provided that the spectrum of $\Gamma$ is shifted such that its smallest eigenvalue equals zero.%--as discussed in Sec. \ref{optimal}.

\section{Geodesics and speed limit}
Let us now consider two points $\rho_1,\rho_2 \in \Q_r$ and the trajectories connecting them. The shortest geodesic is the shortest trajectory with a constant speed---i.e., a constant norm of the tangent vector. To compute it, it is useful to go into a purification space that facilitates a geometric analysis through a fiber bundle construction.
Let us first recall some notions of purifications and fiber bundles. 

\subsection{Purifications, fiber bundle, and geodesics}
Any density matrix $\rho \in \Q_r$ acting on a Hilbert space $\H_\textrm{S}$ with dimension $n$ can be purified using an ancilla modelled on $\H_\textrm{A}$ with $\dim(\H_\textrm{A})\geq r$. Specifically, there always exists a normalized vector $\ket{W} \in \H_\textrm{S} \otimes \H_\textrm{A}$ such that  $\rho=\Tr_\textrm{A}{\ketbra{W}{W}}$. We call such a vector a \textit{purification} of the state $\rho$. It can be explicitly expressed in terms of the spectral decomposition as 
\begin{equation}
    \ket{W} = \sum_{k=1}^r \sqrt{\lambda_k}\ket{\lambda_k}\otimes\ket{a_k}, \label{eq: purification state}
\end{equation}
where $\{\ket{a_k}\}_{k=1}^r$ forms an orthonormal set in $\H_\textrm{A}$. We consider \textit{minimal} purifications by fixing $\dim(\H_\textrm{A})=r$. Even with this restriction, purifications are not unique because of possible local unitary transformations on the ancilla. 

For convenience, we write any, not necessarily normalized, vector $\ket{W} \in \H_\textrm{S} \otimes \H_\textrm{A}$ in the form of a matrix, namely 
\begin{equation}
    \ket{W}\!\!=\!\! \sum_{j,k=1}^r w_{jk} \ket{S_j}\otimes\ket{A_k} \mapsto\!\!\sum_{j,k=1}^r\!\! w_{jk} \ketbra{S_j}{A_k} =W,
\end{equation}
where $w_{jk}\in \CC$, and $\ket{S_j}$ and $\ket{A_k}$ span the basis of $\H_\textrm{S}$ and $\H_\textrm{A}$, respectively. This defines an isomorphism between $\H_\textrm{S} \otimes \H_\textrm{A}$ and the space of linear maps $\B_r := B(\H_\textrm{A},\H_\textrm{S})$. 
Such a matrix representation considerably simplifies computation, as we will see below. In particular we may now write $\rho=\Tr_\textrm{A}{\ketbra{W}{W}} = W W\dg$.

\begin{figure}
     \centering
     \includegraphics[width=1.0\columnwidth]{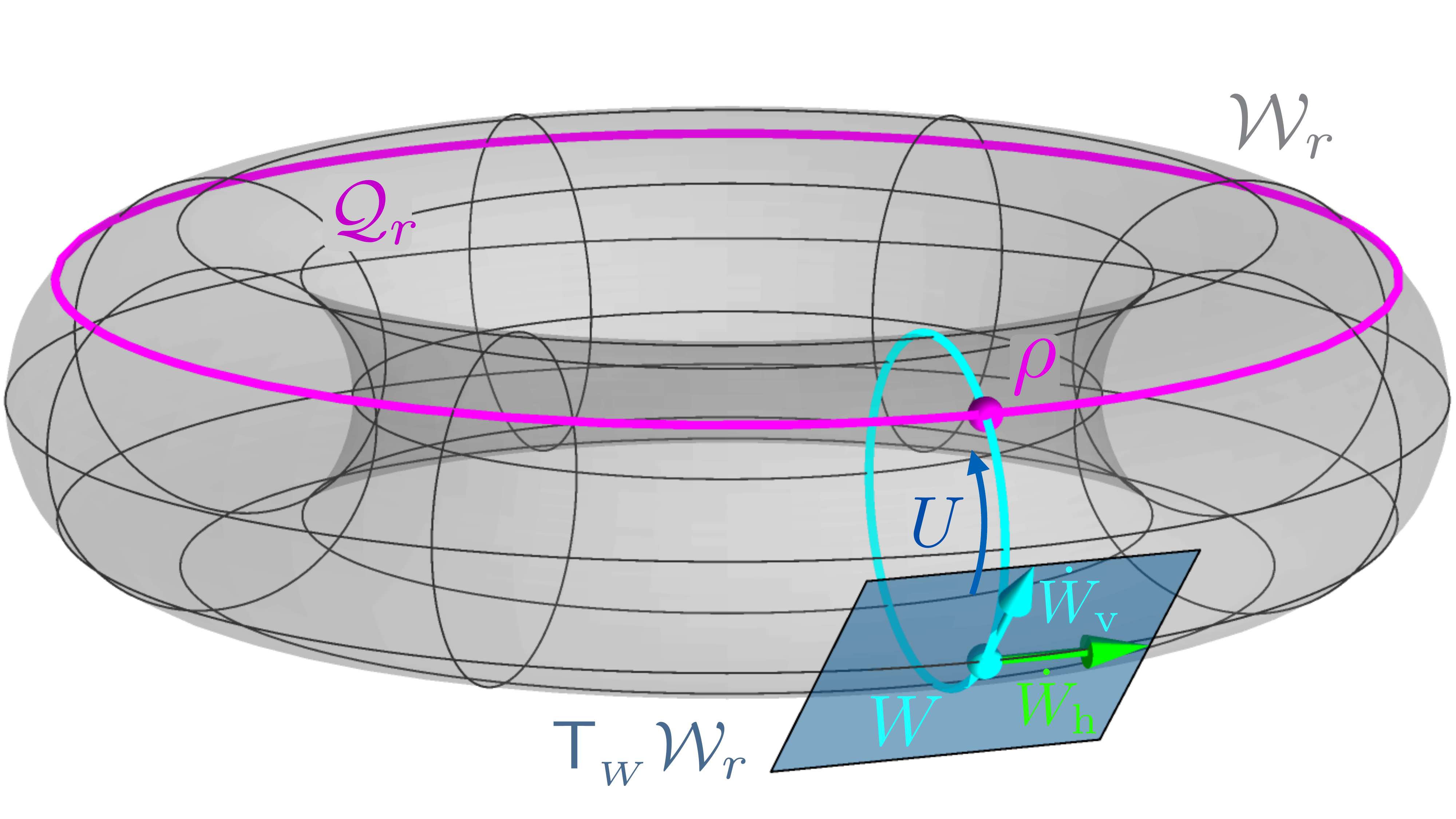}
     \caption{Schematic illustration of the bundle structure (not specific to any $n,r$). To each state $\rho = WW^\dagger$, a fiber $\pi^{-1}(\rho)$ is attached (cyan circle). 
     The unitary group action transforms the purification $W$ to a different point on the fiber $WU$. The vertical space $\mathsf{V}_W\W_r$ forms the tangent space to the fiber while $\mathsf{H}_W\W_r$ forms the orthogonal compliment.}
     \label{fig: bundle}
 \end{figure}
 
We restrict the set $\B_r$ to matrices representing purifications of density matrices of rank $r$ which defines the set
\begin{equation}\label{eq:Wr}
  \!\!\!\! \W_r =\lbrace W  \in \B_r \vert \textrm{Tr} (WW^\dagger)=1 \textrm{ and } \exists (W^\dagger W)^{-1}\rbrace, \!\!\!
\end{equation}
where the first condition corresponds to the normalization and the second is equivalent to having $WW^\dagger$ with a fixed rank $r$. 
In what follows, we refer to the elements of $\W_r$ simply as purifications. 

At each point $\rho \in \Q_r$, we can think of `attaching' the set of all purifications satisfying $\rho = W W\dg$. The triple $(\Q_r, \W_r, \pi)$ forms a \emph{fiber bundle}, with $\pi$ being a projection map 
\begin{equation}
    \pi:\W_r\to\Q_r, \quad \pi(W)=WW^\dagger.
\end{equation}
The pre-image $\pi^{-1}(\rho)$ is known as the fiber over $\rho$ and consists of the purifications for which $WW^\dagger = \rho$. More precisely, it forms a \emph{principal fiber bundle} as one can define a unitary group action on the fibers which account for the unitary freedom on the ancilla system
\begin{equation}
    \Xi:\W^r\times \textrm{U}(r)\to\W^r, \quad\Xi(W,U)=WU.
\end{equation}
See also Appendix~\ref{Appendix: geometry}. 
The projection map, in turn, induces a mapping between the tangent spaces of $\W_r$ and $\Q_r$, namely
\begin{equation}
\begin{split}
\label{differential}
&\mathrm{d}\pi_{_W} : \mathsf{T}_{_W} \W_r \rightarrow \mathsf{T}_{_{WW\dg}} \Q_r, \\
&\mathrm{d}\pi_{_W}(\dot{W}) = \dot{W}W\dg + W \dot{W}\dg.
\end{split}
\end{equation}
The tangent vectors in $\T_{_W} \W_r$ that are tangent to the fiber form the vertical space, defined as $\mathsf{V}_{_W}\W_r = \ker \d\pi_{_W}$.
It may be complemented to form the entire tangent space $\mathsf{T}_{_{W}}\W_r$. The complement is known as the horizontal space, denoted $\mathsf{H}_{_W}\W_r$. We choose this complement to be orthogonal with respect to the metric locally defined as 
\begin{equation}
\begin{split}
\label{euclidean metric}
    &\inner{\cdot}{\cdot}: \mathsf{T}_{_W}\W_r\times\mathsf{T}_{_W}\W_r \to \mathbb{R}, \\
    &\inner{\dot{W}}{\dot{W}'}=\frac{1}{2}\Tr( \dot{W}^\dagger \dot{W}'+\dot{W}'^\dagger \dot{W}),
    \end{split}
\end{equation}
where $\dot{W}$ and $\dot{W}'$ are two tangent vectors at the point $W$. This is a restriction of the Euclidean metric on $\mathsf{T}_{_W} \B_r$.
The horizontal space can then be thought of as orthogonal to the fiber itself; see Fig.~\ref{fig: bundle}. An important property of this metric is that it is invariant with respect to the unitary group action, $\inner{\dot{W}U}{\dot{W}'U} = \inner{\dot{W}}{\dot{W}'}$. This metric is called \textit{right-invariant} and will allow us to connect the metric on $\W_r$ to a metric on $\Q_r$, as seen below.

Any tangent vector $\dot{W}\in\W_r$ can now be decomposed into a vertical and horizontal component $\dot{W} = \dot{W}_\textrm{v} + \dot{W}_\textrm{h}$, with
\begin{subequations}
\label{eq: A and L}
    \begin{align}
    &\dot{W}_\textrm{v} =  W A,\\
    &\dot{W}_\textrm{h} =  LW,
\end{align}
\end{subequations}
where $A\dg =-A$, $L=L\dg$ and $\textrm{Tr}(LWW\dg) = 0$, see for example \cite{hornedal_extensions_2022}.

In addition, the space $\mathsf{H}_{_W}\W_r$ is isomorphic with the tangent space $\mathsf{T}_{_{WW^\dagger}}\Q_r$ \cite{nakahara_geometry_2017}. This, in combination with \eqref{euclidean metric} being right-invariant, allows $\Q_r$ to inherit the metric from $\W_r$. The induced metric turns out to be exactly the Bures metric \eqref{eq:Bures}, namely
\begin{equation}
\label{Bures horizontal}
   (\dot{W}_\textrm{h},\dot{W}_\textrm{h}') = (\dot{\rho},\dot{\rho}')_\textsc{b},
\end{equation}
where $\dot{W}_\textrm{h},\dot{W}_\textrm{h}' \in \mathsf{H}_{_W}\W_r$, $W\in\pi^{-1}(\rho)$, $\d\pi_{_W} \dot{W}_\textrm{h} = \dot{\rho}$ and $\d\pi_{_W} \dot{W}_\textrm{h}' = \dot{\rho}'$. Importantly, the right-invariance of \eqref{euclidean metric} implies that \eqref{Bures horizontal} does not depend on the purification $W$ chosen from the fiber.

Since $\B_r$ can be seen as a real vector space, it can be equipped with an inner product. We choose the inner product compatible with \eqref{euclidean metric}, which takes the same form but with elements from $\B_r$ instead of $\mathsf{T} \W_r$. This inner product allows us to interpret $\textrm{Tr}(W W\dg)=1$ in \eqref{eq:Wr} as a unit sphere. 

The shortest geodesic is the shortest path between two given points parameterized by its arc length. On a sphere, they are the great arcs. The set $\W_r$ is a homogeneous space and forms an open and dense subset of the unit sphere in $\B_r$. As a consequence, the geodesics on $\W_r$ are also great arcs. The geodesic distance between two purifications $W_1, W_2$ is thus the arc length $\arccos \inner{W_1}{W_2}$. Each geodesic in the base space $\Q_r$ can in turn be \textit{lifted} to a great arc in the purification space, i.e., there exists a great arc in $\W_r$ that projects down to the geodesic in $\Q_r$. In particular, the shortest geodesic between $\rho_1$ and $\rho_2$ is a projection of a shortest great arc between the corresponding fibers, i.e., a geodesic between two purifications  $W_1\in\pi^{-1}(\rho_1)$, $W_2\in\pi^{-1}(\rho_2)$ minimizing $\arccos \inner{W_1}{W_2}$ \cite{hornedal_extensions_2022}. This is equivalent to requiring
\begin{equation}\label{eq: positive overlap}
    W_1^\dagger W_2\geq 0,
\end{equation}
as first shown in \cite{uhlmann_transition_1976}; see also Appendix \ref{bures angle}. The corresponding geodesic distance on $\Q_r$ equals $\arccos\Tr\sqrt{W_1^\dagger W_2 W_2^\dagger W_1}$. It may be expressed independently of the purifications as
\begin{equation}
\label{geo dist}
    \dist(\rho_1,\rho_2) = \arccos\sqrt{F(\rho_1,\rho_2)},
\end{equation}
where we recover the Uhlmann fidelity $F(\rho_1,\rho_2) = \left(\Tr\sqrt{\sqrt{\rho_1}\rho_2\sqrt{\rho_1}}\right)^2$ \cite{uhlmann_transition_1976}; see also Appendix \ref{bures angle} for a self-contained derivation.

\begin{figure}
     \centering
     \includegraphics[trim=0 60 0 200,clip,width=1.0\columnwidth]{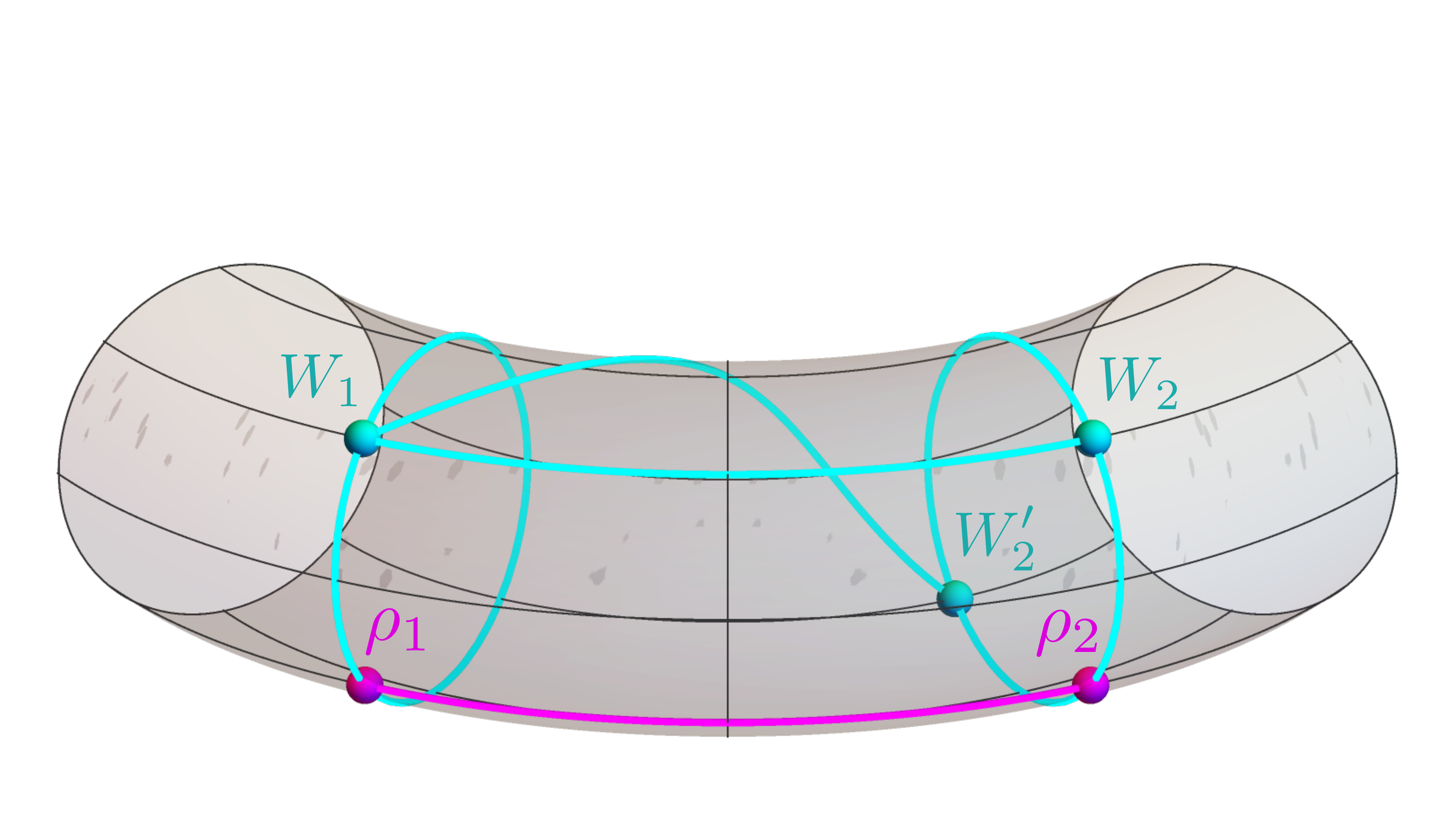}
     \caption{Both paths between $W_1$ and $W_2$, and between $W_1$ and $W_2'$ are geodesic on $\W_r$. The first one however is the shortest geodesic between fibers $\pi^{-1}(\rho_1)$ and $\pi^{-1}(\rho_2)$ and hence its projection onto $\Q_r$ is the shortest geodesic between $\rho_1$ and $\rho_2$.}
     \label{fig: geodesic}
 \end{figure}

\subsection{Speed limit}
\label{speed limit}
A speed limit is defined as a lower bound on the time required to evolve the system from an initial state to a target state under some specified class of dynamics. The Bures metric gives a natural speed limit for general open dynamics in terms of the geodesic distance and the time-averaged speed of the trajectory \cite{uhlmann_energy_1992,frowis_kind_2012, taddei_quantum_2013}. Here, we will describe the speed in terms of a non-Hermitian generator of the evolution and analyze the tightness of the speed limit for different upper bounds of this speed.

A non-Hermitian evolution satisfying \eqref{eq: rho dot H and Gamma} can be lifted to an evolution in the purification space satisfying
\begin{equation}
\label{purified velocity}
    \dot{W} = -iHW - \Tilde{\Gamma}_\rho W,
\end{equation}
where $\Tilde{\Gamma}_\rho$ can be written as $\Gamma - \textrm{Tr}(\Gamma WW\dg)$.
Indeed, the tangent $\dot{\rho} = \mathrm{d}\pi_{_W}(\dot{W})$ can be found using \eqref{differential} and the fact that $\rho = WW\dg$. From \eqref{eq: A and L}, it follows that $\Tilde{\Gamma}_\rho W \in H_{_W}\W_r$, while $iHW$ in general contains both vertical and horizontal contributions. In fact, \eqref{purified velocity} is horizontal iff $\Pi H\Pi = 0$, where $\Pi = \sum_{k:\lambda_k\neq 0} \Pi_k$ is the projection onto the support of $\rho$ \cite{hornedal_extensions_2022}. Consequently, when moving horizontally within the support of the state, the term $\Tilde{\Gamma}_\rho W$ is non-zero, implying that the system is necessarily open. Note that in the special case of a full-rank state, the unitary flow has to be zero, that is, $H=0$, for the evolution to be completely horizontal.

The horizontal part of \eqref{purified velocity} is obtained by decomposing $-iHW$ according to \eqref{eq: A and L},
\begin{equation}\label{eq:Wh}
    \dot{W}_\textrm{h} = LW - \Tilde{\Gamma}_\rho W,
\end{equation}
where  $L$ is an SLD fulfilling $\dot{\rho}_H = L \rho + \rho L$. 
This equation can be found by imposing $d\pi_{_W}(- i H W) = d\pi_{_W}(LW)$ and using \eqref{differential}. By then using the spectral decomposition of $\rho$ and projecting both sides of the equation on $\Pi_j \bullet \Pi_k$, we find 
\begin{equation} \label{Lexplicit}
    L = i\sum_{j,k} \frac{\lambda_j - \lambda_k}{\lambda_j + \lambda_k}\Pi_j H \Pi_k.
\end{equation}

As the metric \eqref{Bures horizontal} on $\mathsf{T}\Q_r$ is defined from the horizontal components of the purifications, it follows that
\begin{equation}
\label{speed 1}
    (\dot{\rho},\dot{\rho})_\textsc{b} = \textrm{Tr}(\dot{W}_\textrm{h}\dg \dot{W}_\textrm{h}) = \Tr((L - \tilde{\Gamma}_\rho)^2\rho) .
\end{equation}
The form of $L$ given in \eqref{Lexplicit} helps to recognize that $\Tr(L^2 \rho) = \frac{1}{4}\mathcal{F}(H,\rho)$, where $\mathcal{F}(H,\rho)$ is the quantum Fisher information of the state with respect to $H$ \cite{petz_state_2008}. 
As a result
\begin{equation}
\label{speed 2}
    (\dot{\rho},\dot{\rho})_\textsc{b} = \frac{1}{4}\mathcal{F}(H,\rho) +\Delta^2 \Gamma + \frac{1}{4}\Tr(\{L,\Gamma\}\rho),
\end{equation}
where $\Delta^2 \Gamma = \Tr(\Gamma^2\rho) - \Tr(\Gamma\rho)^2$.

Combining \eqref{speed 2} and \eqref{geo dist} results in the lower bound of the time to evolve $\rho(0) = \rho_1$ to $\rho(t) = \rho_2$
\begin{equation}
\label{QSL}
    \begin{split}
        t \geq \frac{\arccos{\sqrt{F(\rho_1,\rho_2)}}}{\frac{1}{t}\int_0^t\sqrt{(\dot{\rho},\dot{\rho})_\textsc{b}}\dt'}.
    \end{split}
\end{equation}
We can yet define a weaker speed limit by upper bounding the speed and obtain
\begin{equation}
\label{QSL_tight}
    \begin{split}
        t \geq \frac{\arccos{\sqrt{F(\rho_1,\rho_2)}}}{\frac{1}{t}\int_0^t\sqrt{\Delta^2 H +\Delta^2 \Gamma -i\Tr([H,\Gamma]\rho)}\dt'}.
    \end{split}
\end{equation}
To see this, note that the integrand in the denominator of \eqref{QSL_tight} is equal to $\sqrt{(\dot{W}',\dot{W}')}$, where $\dot{W}' = -i\Tilde{H}_\rho W - \Tilde{\Gamma}_\rho W$. 
%\oap{\sout{is also a lift of $\dot{\rho}$, i.e., $\d\pi_{_W}(\dot{W}')=\dot{\rho}$}}. 
As before, the tilde with subscript $\rho$ indicates a shift of the operator such that the expectation value with respect to $\rho$ equals zero. The vector $\dot{W}'$ is also a lift of $\dot{\rho}$, i.e., $\d\pi_{_W}(\dot{W}')=\dot{\rho}$ and hence its horizontal component is equal to $\dot{W}_\mathrm{h}$. The inequality \eqref{QSL_tight} then follows from $(\dot{\rho}, \dot{\rho})_\textsc{b} = (\dot{W}_\textrm{h}',\dot{W}_\textrm{h}') \leq (\dot{W}',\dot{W}')$. This is saturated iff $\Pi \Tilde{H}_\rho \Pi = 0$, which is equivalent to $\Pi H\Pi = \alpha\Pi$, $\alpha\in\mathbb{R}$. It holds trivially for pure states but not for general mixed states \cite{hornedal_extensions_2022}.

In practice, the time-averaged speed needs to be replaced with a constant that upper bounds it. In this way, a lower bound on the evolution time can be computed without the need to solve for the whole evolution. A similar speed limit to \eqref{QSL} has been proposed in \cite{thakuria_generalised_2023}, which holds in the space of purifications but may be violated for density matrices if one does not choose the correct purifications. More precisely, the derivation presented therein relies on using the metric that \eqref{euclidean metric} induces on the projectivization of $\W_r$ ($\mathcal{PW}_r$), i.e., two purifications are identified as the same element if they differ by a phase. The distance between two elements in $\mathcal{PW}_r$ represented by the purifications $W_1$ and $W_2$ is then given by the Fubini-Study distance $\arccos{\abs*{\Tr W_1^\dagger W_2}}$. However, it may be problematic to apply this speed limit to mixed states. More precisely, upper bounding the speed in $\mathcal{PW}_r$ by some constant $\omega$ will not guarantee that $\big(\arccos{\abs*{\Tr W_1^\dagger W_2}}\big)/ \omega$ will lower bound the evolution time from $\rho_1 = W_1W_1^\dagger$ to $\rho_2 = W_2W_2^\dagger$. Indeed, consider the case when $W_1^\dagger W_2$ does not satisfy \eqref{eq: positive overlap}. We can then find a different purification $W_2'$ satisfying $W_2'W_2'^\dagger = \rho_2$ and $W_1^\dagger W_2' \geq 0$. There exists an evolution $W_t$ that connects $W_1$ and $W_2'$ with length $\arccos{\abs*{\Tr W_1^\dagger W_2'}}$ and that evolves at a constant speed equal to $\omega$. The projection of $W_t$ then yields an evolution between $\rho_1$ and $\rho_2$ that respects the upper bound on the speed $\omega$ but with an evolution time strictly smaller than $\big(\arccos{\abs*{\Tr W_1^\dagger W_2}}\big)/ \omega$. 
The bounds \eqref{QSL} and \eqref{QSL_tight} avoids this issue since they are derived through a metric well defined on $\Q_r$, rather than on $\mathcal{PW}_r$.

%Consider a non-Hermitian generator $K$ evolving $W_1$ to $W_2$ which saturates the speed limit in \cite{thakuria_generalised_2023}, $\tau_W$, and let $\V_\tau$ denote the time-averaged speed of the trajectory with respect to the metric in $\mathcal{PW}_r$. Whenever $W^\dagger_0W_\tau$ does not satisfy Eq. \eqref{eq: positive overlap}, there exist another Hamiltonian $K'$ evolving $\rho_1$ to $\rho_2$ in a time strictly smaller than $\tau_W$ but with the same time-averaged speed $\V_\tau$ in $\mathcal{PW}_r$.

\subsection{Generator of geodesic}
We move now to derive an explicit expression for the shortest geodesics in $\Q_r$ and describe a non-Hermitian dynamics that can generate it. This section is based on \cite{ericsson_geodesics_2005, spehner_bures_2023} and extends the results to states with rank $r<n$.

The shortest geodesic in $\Q_r$ between two states $\rho_1$ and $\rho_2$ can be lifted to the shortest geodesic between the fibers in $\W_r$ over these states, that is, to a geodesic between $W_1,W_2$ satisfying the condition \eqref{eq: positive overlap}. As an arc of a great circle, it takes the form
\begin{equation}
    W(\tau) = \frac{1}{\sin \theta}\left(\sin(\theta-\tau)W_1 + \sin \tau W_2\right), \label{eq: W geodesic}
\end{equation}
where $\theta = \arccos{\inner{W_1}{W_2}}$ is the geodesic distance. The geodesic on $\Q_r$ is then given by $\rho_g(\tau)=W(\tau)W^\dagger(\tau)$.

We now ask what evolution follows this path. More precisely, we look for a geodesic evolution operator $G_\textrm{g}(\tau)$ such that $W(\tau)=G_\textrm{g}(\tau)W_1$, or equivalently $\rho_g(\tau)=G_\textrm{g}(\tau)\rho_1G_\textrm{g}^\dagger(\tau)$. To do so, we introduce the operator $M$ that maps the initial purification onto the final one, $W_2 = M W_1$. This operator is defined uniquely only for full-rank states. For $r<n$, we have some freedom and pick
\begin{multline}
\label{M}
        M :=  W_2W_1^+ + iR \\
         +(\id {-} \Pi_1)W_2W_1^+ + (W_2W_1^+)^\dagger(\id {-} \Pi_1 ), 
\end{multline}
where $ \Pi_1=  W_1 W^+_1 $ is the projection onto the support of $W_1$, defined from the Moore-Penrose inverse $W_1^+$.
We choose $R\geq 0$ such that $R=(\id-\Pi_1)R \, (\id-\Pi_1)$ to ensure invertibility of the geodesic evolution operator $G_\textrm{g}(\tau)$, and detail this choice in Appendix \ref{app:chi}.

Any point $W(\tau)$ on the shortest trajectory can thus be obtained from the initial state as $W(\tau)=G_\textrm{g}(\tau) W_1$, where
\begin{equation}
\label{X}
    G_\textrm{g}(\tau) = \frac{1}{\sin \theta}\left(\sin(\theta-\tau)\mathds{1} + \sin \tau \,M\right).
\end{equation}
Trivially, conjugation with such an operator yields the evolution on $\Q_r$.

Note that such dynamics preserves the trace along the chosen geodesic but not necessarily along another trajectory. We therefore define the map
\begin{equation}
  \!\!\!  \Phi_{\tau}^\textrm{g}:\Q_r\to\Q_r,\:\:\:\:\: \Phi_{\tau}^\textrm{g}(\rho):=\frac{G_\textrm{g}(\tau)\rho G_\textrm{g}^\dagger(\tau)}{\Tr(G_\textrm{g}(\tau)\rho G_\textrm{g}^\dagger(\tau))},\!\!
\end{equation}
which is of the form \eqref{eq: normalised map}. As described in Sec. \ref{Non-Hermitian dynamics and fixed-rank manifold}, the generator of this evolution is given by
\begin{equation}
\label{geodesic generator}
    K_\textrm{g}(\tau) := i\dv{\tau}G_\textrm{g}(\tau) \,G_\textrm{g}^{-1}(\tau).
\end{equation}
Note that, by construction,  the trajectory in the purification space is horizontal, i.e., $\dot{W}(\tau) \in H_{W(\tau)} \W_r$. For full-rank states, this implies that $K_\textrm{g}(\tau)$ is anti-Hermitian at all times and corresponds to the unique solution of $K_\textrm{g}(\tau)\rho_\textrm{g}(\tau) + \rho_\textrm{g}(\tau)K_\textrm{g}(\tau) = i\dv{\tau} \rho_\textrm{g}(\tau)$.

In general, the generator \eqref{geodesic generator} is time-dependent and not unique. It prescribes the motion along the shortest path between two states with constant speed. A natural question is whether it can be made time-independent by possibly loosening the restriction on the speed being constant. In the qubit case, it is always possible to do so. The resulting generator is necessarily anti-Hermitian for the full-rank state, as illustrated in the application below.
In the general case, however, a time-independent generator does not always exist. This claim is supported by the counter-example of a qutrit, which we detail in Appendix \ref{sec:qutrit}. 

\begin{figure*}
     \centering
     \includegraphics[width=0.9\textwidth]{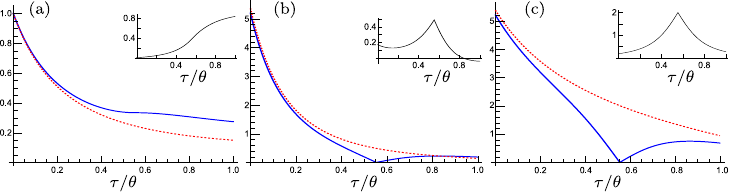}
     \caption{(a) State norm $\Tr(\rho)$ along the \emph{un-normalized} evolution between $\rho_1$ and $\rho_2$ generated by \eqref{geodesic generator} (red dashed) and its optimization  (solid blue). Inset: relative difference of the norms.
     (b;c) Decay rate $\gamma$ for the generator \eqref{geodesic generator} (red dashed) and $\gamma_\textrm{opt}$ for its optimization (solid blue) along the (b) \emph{un-normalized} and (c) \emph{normalized} evolution between $\rho_1$ and $\rho_2$. Inset: difference of the rates.}
     \label{fig:qubit}
 \end{figure*}

\section{Application: Non-Hermitian control in a qubit \label{qubit}}
To illustrate our findings, we consider the shortest geodesic between the qubit states $\rho_1= \frac{1}{2}(\id + \frac{1}{5}\sigma_z)$ %\mqty( 0.9 & 0 \\ 0 & 0.1 )
and $\rho_2=\frac{1}{2}(\id -\frac{1}{2}\sigma_x- \frac{1}{2}\sigma_z)$. %\mqty( 0.25 & -0.25 \\ -0.25 & 0.75 ).
This trajectory is generated using \eqref{geodesic generator} and then compared with the optimization described in Sec.~\ref{optimal}. For a fair comparison, the spectrum of the anti-Hermitian part is shifted so that the smallest eigenvalue equals zero for both generators. 
Fig.~\ref{fig:qubit} shows how the state norm and its decay rate change over time for the two trajectories. 
As expected, we see that the optimization leads to a smaller decay of the norm.
Notably, the relative difference in norms---indicating how many more times the experiment must be re-run for the trajectory $\rho$ to achieve the same success rate as the optimized trajectory $\rho_\textrm{opt}$---approaches one.

Figure ~\ref{fig:qubit} also depicts the decay rate $\gamma$ calculated for both un-normalized and normalized states along the evolution. The former is simply the derivative of state's norm and can thus be seen as a ``global" decay rate. From the continuous measurement point of view, it describes the behavior of the whole ensemble of experiments. The latter is renormalized by decreasing trace and hence corresponds to a ``local" decay of the state. For a given experiment, it describes the probability per unit time of not surviving the measurement, conditioned on the state having survived this long. It may serve as an illustration of the inequality \eqref{eq: optimal Gamma}.
The decay rate for the optimized dynamics $\gamma_\textrm{opt}$ drops to zero and becomes non-analytic at the time when the state passes the minor axis of an ellipse associated with the geodesic, as discussed below. At that point, such an ellipse is tangent to a concentric sphere, meaning that the velocity at this point has no commutative component and can be fully expressed in terms of a Hamiltonian vector field. That is exactly what the optimization procedure does. The non-analyticity there is caused by the level crossing in $\Gamma_\textrm{c}$.

In order to construct these generators, we first took the trivial purifications of the states, respectively $W_1=\sqrt{\rho_1}$ and $W_2'=\sqrt{\rho_2}$, which in general do not satisfy the condition \eqref{eq: positive overlap}. Keeping $W_1$ fixed and using the unitary freedom, we then found a purification $W_2$ satisfying \eqref{eq: positive overlap} by picking $W_2=W_2'U^\dagger$, where $W_1^\dagger W_2'=P U$ is the polar decomposition with $P\geq 0$ and $U$ unitary. We defined $W(\tau)$ according to \eqref{eq: W geodesic} and found $G_g(\tau)=W(\tau)W_1^{-1}$ (notice the simplification due to the existence of the inverse $W_1^{-1}$). Finally, we defined $K_g(\tau)$ according to \eqref{geodesic generator}.

\begin{figure}
     \centering
     \includegraphics[trim=230 17 230 0,clip,width=.80\columnwidth]{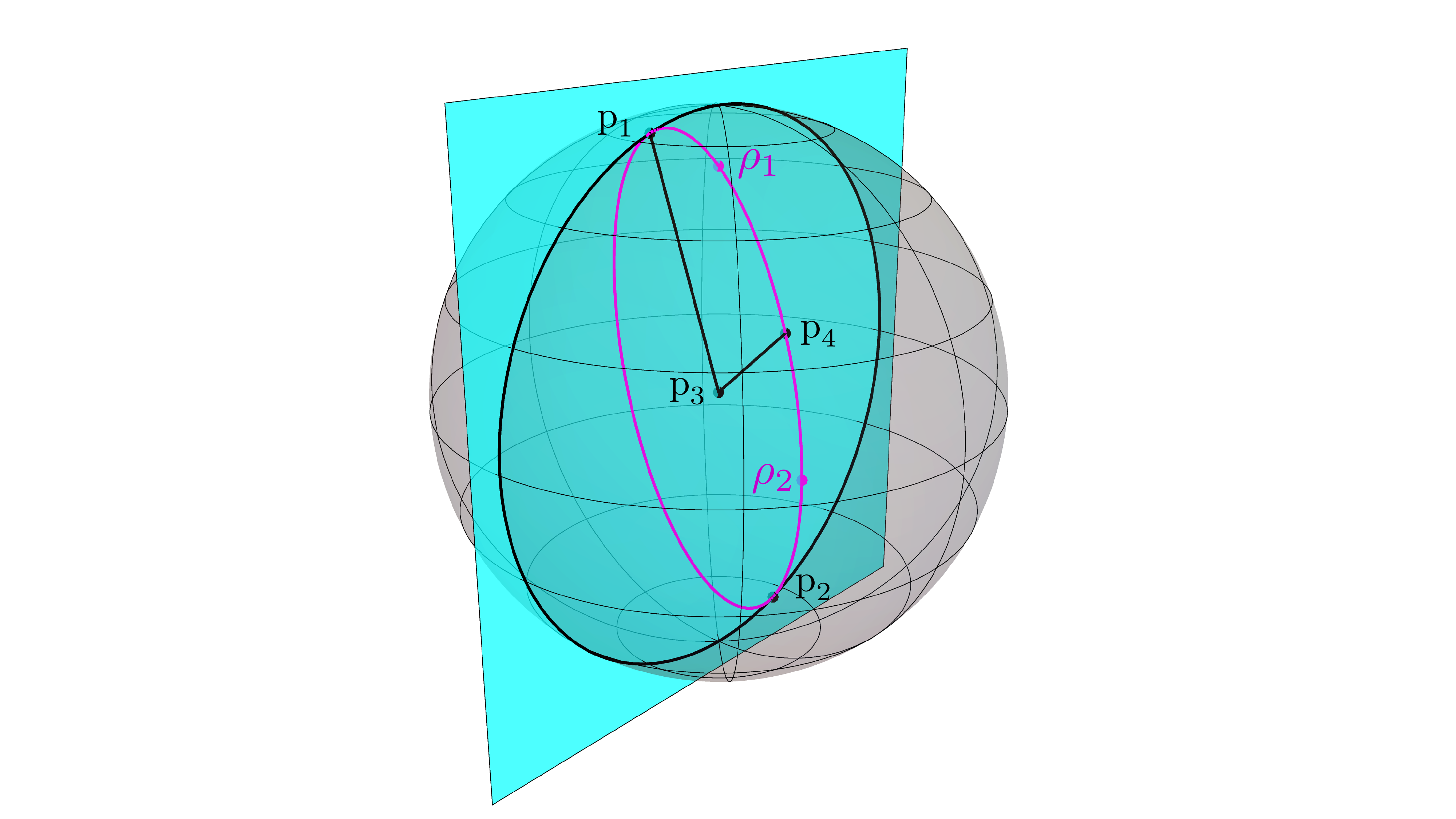}
     \caption{The figure depicts the plane (cyan) containing the maximally mixed state ($\textrm{p}_3$) and the initial and final states ($\rho_1$ and $\rho_2$). The intersection with the Bloch sphere forms a great circle, and the ellipse (magenta) for which the shortest geodesic is a segment of is the unique ellipse intersecting $\rho_1$ and $\rho_2$ and with its two points on its major axis ($\textrm{p}_1$ and $\textrm{p}_2$) lying on the great circle. The point $\textrm{p}_4$ lies on the minor axis of the ellipse and is always an interior point for full-rank states.}
     \label{fig: qubit geodesic}
 \end{figure}

Interestingly, the image of any shortest geodesic in a qubit can be generated by time-independent anti-Hermitian Hamiltonians (see Appendix \ref{time-independent generators} for details) being a rescaled generator \eqref{geodesic generator}. Consequently, each shortest geodesic will form a segment of an ellipse that has its centre at the fully mixed state and the points contained along its major axis being the two eigenstates of \eqref{geodesic generator}, as illustrated in Fig.~\ref{fig: qubit geodesic}.

\section{Conclusion}
\label{section: conclusion}
Starting from the basic description of quantum evolution, we have shown how a trace-preserving non-Hermitian evolution preserves the rank of the density matrix. This allowed us to develop a geometrical understanding of such dynamics. Specifically, the tangent vector (master equations \eqref{eq: rho dot} and \eqref{eq: rho dot H and Gamma}) can be decomposed in the instantaneous state eigenbasis into a direction that generates coherence and another that changes population only. These directions are mutually orthogonal. The incoherent part can further be decomposed into a component that lifts degeneracy and decreases the von Neumann entropy and a part that can be viewed as classical since it does not change the eigenbasis. By decomposition of the anti-Hermitian part of the generator into the coherent and incoherent direction, we showed that the non-Hermitian dynamics that maximizes the success rate of a given trajectory is one whose anti-Hermitian part commutes with the instantaneous state.

Using techniques of purification, we then analyzed the tightness of a quantum speed limit \eqref{QSL_tight}, which is saturated when the evolution moves along the shortest geodesic linking the initial and final state. An explicit expression for this geodesic was derived and used to construct a generator of it \eqref{X}, that is, a non-Hermitian evolution that saturates the speed limit. 

It is natural to ask whether the above geometrical interpretation could be extended to rank-changing dynamics, as resulting from a Lindblad master equation, including the jump terms---see e.g., \cite{ciaglia_dynamical_2017, carinena_tensorial_2017, ciaglia_differential_2017,viennot_purification_2018}  for different approaches.
Although the set of all quantum states $\Q$ is no longer a manifold, 
it forms a Whitney stratified space, that is, a collection of manifolds glued together, and as such, it can be endowed with a generalized differential structure \cite{andrea_pseudo_2021}.
Then, it is possible to extend the notion of tangent spaces to $\Q$ and distinguish the commutative and unitary directions. The precise treatment of this problem exceeds the scope of this paper.

We hope that the insights presented in this manuscript and the intuition that can be gained from our results will help develop new control protocols in quantum systems.

\section{Acknowledgements}

The authors thank Dan Allan for early contributions and Ole Sönnerborn and Pablo Martínez-Azcona for providing valuable feedback on the manuscript. This research was funded by the Luxembourg National Research Fund (FNR, Attract grant QOMPET 15382998, and grant 17132054).

\appendix

\section{Geometrical preliminaries}
\label{Appendix: geometry}
This Appendix provides a brief overview of the concepts from differential geometry used in the main text.

\subsection{Differentiable manifolds}
Let us start by reviewing the notion of a manifold. A real $d$-dimensional manifold $\M$ is a topological space locally homeomorphic to open subsets of $\mathbb{R}^d$. For example, a sphere is a $2$-dimensional manifold. However, a sphere with an attached string is not because the string itself is $1$-dimensional. If these local homeomorphisms, called charts, ``glue together" in a differentiable way, the local similarity to real space allows for introducing the notion of differentiability, otherwise not defined on the topological spaces. We call the collection of such charts covering the whole $\M$ a \emph{differentiable atlas} and the manifold itself a \emph{differentiable manifold}.

One may ask if such a differentiable manifold could not be embedded in $\mathbb{R}^n$ with $n\geq d$ and inherit the notion of differentiability from there instead of having it defined locally in the charts. It turns out that it is indeed possible, as made precise by the Whitney embedding theorem, and that the two notions are compatible (in the sense that every embedding induces a differentiable atlas and every differentiable atlas defines an embedding). Throughout this paper, we use the later picture as a natural embedding of our manifolds exists in higher dimensional spaces. However, all notions discussed here can be seen as internal to the manifold without the need for any external space, which is exactly where the true power of the geometrical picture lies.

\subsection{Tangent spaces and bundles}
The tangent space is one of the most important and used notions in differential geometry. Consider a differentiable path $\gamma:[0,1]\to \M$ on a differentiable $d$-dimensional manifold $\M$ (embedded in $\mathbb{R}^n$) passing through a point $p$. Trivially, the velocity of $\gamma$ at $p$ is given by its derivative at $p$ (in the embedding) and is tangent to $\gamma$. The set of all possible velocities of all differentiable paths passing through $p$ is the \emph{tangent space} $\mathsf{T}_p \M$. In fact, $\mathsf{T}_p \M$ can be thought of as a $d$-dimensional hyperplane tangential to $\M$ at point $p$.

The velocity of a curve at each of its points belongs to different tangent spaces. It would be useful to have a unified description of this velocity along the whole curve. The proper way to do so is to introduce the notion of a tangent bundle as a locally trivial fiber bundle. 

A locally trivial fiber bundle, later called just fiber bundle, is a triple $(\E,\pi,\M)$ consisting of a total space $\E$ and a base space $\M$, both being differentiable manifolds and a surjective map $\pi:\E\to\M$, called projection, satisfying the \emph{local triviality condition}. Namely, $\E$ has to look locally like a product space $\M\times\F$ for some manifold $\F$ called the \emph{fiber}. More formally, we require that there exist a smooth manifold $\F$, such that for any $p\in\M$ there exists a neighborhood $U$ and a smooth homeomorphism $\phi$ such that the following diagram commutes
\begin{equation} 
    \begin{tikzcd}
        \E \supset \pi^{-1}(U) \arrow[d,"\pi"] \arrow[r,"\phi"] &  U\times\F \arrow[dl,"\textrm{proj}_1"]\\
         U &  
    \end{tikzcd}.
\end{equation}

The \emph{tangent bundle} $\mathsf{T}\M$ is then an example of a fiber bundle over $\M$ with the total space being a disjoint union of all tangent spaces $\E = \bigsqcup_{p\in\M} \mathsf{T}_p\M$ and a natural projection $\mathsf{T}\M \supset \mathsf{T}_p\M \ni X \mapsto \pi(X) = p$. A velocity of $\gamma$ is then a path in $\mathsf{T}\M$. However, we are still unable to talk about its change along the path as we did not specify how to relate different fibers $\mathsf{T}_p\M\subset \mathsf{T}\M$ and take derivatives. We are going to fill this gap in the next subsection.

We can further enrich the fiber bundle by adding a smooth right action of a Lie group $G$ on the total space $\E$, which preserves the fibers and acts freely and transitively on them. Such a fiber bundle is then called a principal fiber bundle.

\subsection{Connections, vertical and horizontal spaces}
To connect different fibers, we need to add yet another structure, a \emph{connection}. A connection is a smooth way of splitting all tangent spaces to the total space $\mathsf{T}_X\E$ into two complementary subspaces: a \emph{vertical space} $\mathsf{V}_X\E$ consisting of directions along the fiber $\F$, and a \emph{horizontal space} $\mathsf{H}_X\E$. Their disjoint unions together with the bundle projection $\pi':\mathsf{T}\E\to \E$ form vertical $\mathsf{V}\E$ and horizontal $\mathsf{H}\E$ bundles respectively. 

Note that specifically, to talk about a change in the velocity of a curve or acceleration, we have to define connection on a tangent bundle with a tangent bundle as its base space $\mathsf{T}\mathsf{T}\M$.

\subsection{Riemannian manifolds}
So far, we have defined a differentiable manifold and a tangent bundle over it. It is easy to check that every $\mathsf{T}_p\M$ is a vector space. However, we cannot yet quantify how different are the vectors, i.e., there is no scalar product. It is tempting just to inherit the Euclidean inner product from the embedding (this is what we indeed do in the main text, as the considered embeddings will be, in some sense, natural). In general, however, a different choice may be more natural (although Nash's theorem states that for any such choice, there exists an embedding inducing the same local inner product). A smooth assignment of scalar products to each tangent space is called a Riemannian metric, and a differentiable manifold now endowed with such a metric becomes a Riemannian manifold.
\vspace{1cm}

As an additional side note, let us mention that there exist complex manifolds that locally look like $\mathbb{C}^d$ and not $\mathbb{R}^d$. For example, a physical Hilbert space is, in a natural way, such a complex manifold. Complex manifolds can, however, be seen as real manifolds of double the dimension with an additional structure, a particular endomorphism of the tangent bundle, which amounts to specifying the multiplication by an imaginary unit. We use this fact to introduce a natural Riemannian structure on Hilbert spaces.

\section{Fixed rank dynamics}
\subsection{Global success rate optimization}\label{app: opti}
Consider a generic non-Hermitian Hamiltonian $K = H -i\Gamma$. For a given initial state $\rho_0$, it generates the un-normalized dynamics $[0,1]\ni t\mapsto\rho(t)$. 
The dynamics reaches the final state with probability $\Tr \rho(1)$.
Our goal is to construct a non-Hermitian Hamiltonian $K'=H'-i\Gamma'$ such that $\rho'(t)$ is proportional to $\rho(t)$ at all times, i.e., the normalized evolutions of $\rho_0$ agree, and moreover, the probability of reaching the final state $\Tr\rho'(1)$ is the maximal possible.

The Hamiltonian $K_\textrm{opt}$ constructed as in Section \ref{optimal} satisfies these demands. In fact, suppose that there exists $K'\neq K_\textrm{opt}$ such that $\Tr \rho'(1) > \Tr\rho_\textrm{opt}(1)$. Initially, the decay rate is lower bounded $\Tr(\Gamma'(0) \rho_0) \geq \Tr(\Gamma_\textrm{opt}(0)\rho_0)$. By continuity, there exists a time $\tau<1$ such that $\Tr\rho'(\tau) = \Tr\rho_\textrm{opt}(\tau)$ and thus $\rho'(\tau) = \rho_\textrm{opt}(\tau)$. At that point, however, the decay rate is again bounded by the locally optimal one $\Tr(\Gamma'(\tau) \rho'(\tau)) \geq \Tr(\Gamma_\textrm{opt}(\tau)\rho_\textrm{opt}(\tau))$. It implies that $\Tr\rho'(t) \leq \Tr\rho_\textrm{opt}(t)$ at all times, contradicting our assumption that $\Tr \rho'(1) > \Tr\rho_\textrm{opt}(1)$. Therefore, the local optimization is also a global optimization.

\subsection{Orthogonality between coherent and incoherent directions}
\label{app: orth}

The inner product $\langle\cdot,\cdot\rangle$ corresponding to the monotone metric \eqref{eq: monotone metric} can be computed through the polarization identity
\begin{equation}
\label{polarization}
    \langle\dot{\rho},\dot{\rho}'\rangle = \frac{1}{4}\big(\norm{\dot{\rho}+\dot{\rho}'}^2 - \norm{\dot{\rho}-\dot{\rho}'}^2\big).
\end{equation}
For any pair of tangent vectors $\dot{\rho}_\textrm{u} \in \mathsf{U}_\rho\,\mathcal{Q}_n$ and $\dot{\rho}_\textrm{c} \in \mathsf{C}_\rho\,\mathcal{Q}_n$, we can choose an eigenbasis common to $\rho$ and $\dot{\rho}_\textrm{c}$ such that $(\dot{\rho}_\textrm{c})_{jk} = 0$ $\forall j\neq k$ and $(\dot{\rho}_\textrm{u})_{jj} = 0$ $\forall j$. From \eqref{eq: monotone metric}, we have
\begin{equation}
\label{pythagoras}
    \begin{split}
        &\norm{\dot{\rho}_\textrm{c}\pm\dot{\rho}_\textrm{u}}^2 = \sum_{j,k}\abs{(\dot{\rho}_\textrm{c}\pm \dot{\rho}_\textrm{u})_{jk}}^2 c(\lambda_j,\lambda_k) \\
        &=\sum_{j}\abs{(\dot{\rho}_\textrm{c})_{jj}}^2 c(\lambda_j,\lambda_j) + \sum_{j\neq k}\abs{(\dot{\rho}_\textrm{u})_{jk}}^2 c(\lambda_j,\lambda_k) \\
        &=\norm{\dot{\rho}_\textrm{c}}^2 + \norm{\dot{\rho}_\textrm{u}}^2.
    \end{split}
\end{equation}
It then follows from \eqref{polarization} that $\langle\dot{\rho}_\textrm{u},\dot{\rho}_\textrm{c}\rangle = 0$.

To show orthogonality between two tangent vectors $\dot{\rho}_\textrm{cl} \in \mathsf{Class}_\rho\,\mathcal{Q}_n$ and $\dot{\rho}_\textrm{l} \in \mathsf{L}_\rho\,\mathcal{Q}_n$, consider a common eigenbasis between $\rho$, $\dot{\rho}_\textrm{cl}$ and $\dot{\rho}_\textrm{l}$; $\{\ket{\lambda_{jk}}\}$, where the first index runs over the number of distinct eigenvalues and $k$ over the multiplicities $m_j$. We let $\lambda_j$ denote the distinct eigenvalues so that $\lambda_{jk} = \lambda_j$ $\forall k\in\{1,2,\dots m_j\}$. We can then write $\dot{\rho}_\textrm{cl} = \sum_j\sum_{k=1}^{m_j} \mu_j \ketbra{\lambda_{jk}}$ and $\dot{\rho}_\textrm{l} = \sum_j\sum_{k=1}^{m_j} \nu_{jk} \ketbra{\lambda_{jk}}$, where $\sum_{k=1}^{m_j}\nu_{jk} = 0\impliedby \Tr(\dot{\rho}_\textrm{l}\Pi_j) = 0$. We then have
\begin{equation}
    \begin{split}
        &\norm{\dot{\rho}_\textrm{cl}\pm \dot{\rho}_\textrm{l}}^2 =\sum_j\sum_{k=1}^{m_j}\abs{(\dot{\rho}_\textrm{cl}\pm \dot{\rho}_\textrm{l})_{jk,jk}}^2c(\lambda_{jk},\lambda_{jk})\\
        &= \sum_jc(\lambda_{j},\lambda_{j}) \sum_{k=1}^{m_j}(\mu_j\pm \nu_{jk})^2\\
        &= \sum_jc(\lambda_{j},\lambda_{j}) \sum_{k=1}^{m_j}(\mu_j^2 + \nu_{jk}^2).
    \end{split}
\end{equation}
Together with \eqref{polarization}, it follows that $\langle\dot{\rho}_\textrm{cl},\dot{\rho}_\textrm{l}\rangle = 0$.

In conclusion, we have shown that the subspaces $\mathsf{U}_\rho \mathcal{Q}_n$, $\mathsf{Class}_\rho \mathcal{Q}_n$ and $\mathsf{L}_\rho \mathcal{Q}_n$ are mutually orthogonal to one another with respect to any monotone metric.

\subsection{The Bures angle}
\label{bures angle}

The Bures angle between two states $\rho_1$ and $\rho_2$ equals the length of the shortest spherical arc connecting the respective fibers and reads
\begin{equation}
    \Theta(\rho_1,\rho_2) = \inf_{W_1,W_2} \arccos(W_1,W_2),
\end{equation}
where $(W_1,W_2)=\Re \Tr (W_1^\dagger W_2)$ is the euclidean inner product. 
The minimization over the purification can be done by fixing purifications $W_1$, $W_2'$ of $\rho_1$, $\rho_2$ respectively, and then considering all possible gauge transformations $W_2 = W_2'U$. The angle is minimized when the inner product between $W_1$ and $W_2$ is maximized.

Let $A$ be an operator and let $\lbrace\ket{k}\rbrace$ be the set of the eigenvectors of $AA\dg$. We then have
$\Re(\Tr A )\leq \sum_k \abs{\bra{k}\!A\!\ket{k}} =\sum_{k} \sqrt{\bra{k}\!A\!\ket{k}\!\bra{k}\!A\dg\!\ket{k}} \leq \sum_k \sqrt{\bra{k}\!AA\dg\ket{k}} =\Tr\sqrt{AA\dg}$. The inequalities are saturated iff $A$ is positive. If we now take $A=W^\dagger_1W_2$ we see that the upper bound is independent of $U$ and the angle is minimized iff $W^\dagger_1W_2 \geq 0$.

Note that $A\geq 0 \iff A = \sqrt{AA^\dagger} = \sqrt{W_1^\dagger W_2W_2^\dagger W_1} = \sqrt{\sqrt{\rho_1} \rho_2 \sqrt{\rho_1}}$ where we fixed $W_1 = \sqrt{\rho_1}$ without loss of generality. Consequently 
\begin{equation}
    \Theta(\rho_1,\rho_2) = \arccos{\Tr \sqrt{\sqrt{\rho_1} \rho_2 \sqrt{\rho_1}}},
\end{equation}
which recovers the Uhlmann fidelity. 

\subsection{Invertibility of the geodesic evolution operator \label{app:chi}}
We show that the evolution operator $G_\textrm{g}(\tau)$ \eqref{X} has a trivial kernel and is thus invertible for all $\tau \in [0,\theta]$.

It is enough to ensure that there are no negative or zero eigenvalues in the spectrum of $M$. Indeed, the condition $\forall \ket{\psi}\in\H: M\ket{\psi} \neq -\frac{\sin(\theta-\tau)}{\sin{\tau}}\ket{\psi} $ is equivalent with the kernel of $G_\textrm{g}(\tau)$ being trivial. 
%\sout{We construct $M$ with complex eigenvalues to ensure no negative eigenvalues exist. }

Using the projectors $W_1W_1^+$ and $(\id - W_1W_1^+)$, the operator $M$ can be expressed as the symmetric block matrix
\begin{equation}
    M = \begin{pmatrix}
        A&C^\dagger\\
        C&B
    \end{pmatrix}, \quad A\geq 0,
\end{equation}
where dim$(A)=r$ and dim$(B)=n-r$. 
We let this matrix act on any vector
\begin{equation}
    \begin{pmatrix}
        A&C^\dagger\\
        C&B
    \end{pmatrix}\begin{pmatrix}
        a\\
        b
    \end{pmatrix} = \begin{pmatrix}
        Aa + C^\dagger b\\
        Ca + Bb
    \end{pmatrix}, 
\end{equation}
and compute the inner product between the vector and its image with respect to $M$. It reads $(a,b)_M:=\inner{a}{Aa} + \inner{b}{Bb} + 2\Re\inner{b}{Ca}.$
Now, choosing $B$ such that $B = \pm iR$ where $R>0$ is sufficient for $(a,b)_M$ to never be a negative number. This implies that no negative numbers are included in the spectrum of $M$, and thus ensures $G_g(\tau)$ has a trivial kernel.

\subsection{Time-independent geodesic generator --  a counterexample \label{sec:qutrit}}
In the main text, we derive the generator of the geodesic \eqref{geodesic generator}, which, in general, is not unique and is time-dependent. We asked whether it could be made time-independent by loosening the constant speed requirement. 
Here, we show an example in a qutrit for which this is not possible. 
Consider the initial and target states 
\begin{equation}
    \rho_1 = \begin{pmatrix}
        \frac{1}{5}&0&0\\
        0&\frac{2}{5}&0\\
        0&0&\frac{2}{5}
    \end{pmatrix},\quad \rho_2 = \begin{pmatrix}
        \frac{1}{5}&0&0\\
        0&\frac{3}{10}&0\\
        0&0&\frac{1}{2}
    \end{pmatrix}.
\end{equation}
In this case, choosing the purifications $W_1 = \sqrt{\rho_1}$ and $W_2 = \sqrt{\rho_2}$ satisfies \eqref{eq: positive overlap} and, according to \eqref{M}-\eqref{X} and \eqref{geodesic generator}, the generator of the evolution at initial time reads
\begin{equation}
    K_\textrm{g}(0) = i\begin{pmatrix}
        \frac{1-\cos{\theta}}{\sin{\theta}}&0&0\\
        0&\frac{\frac{\sqrt{3}}{2}-\cos{\theta}}{\sin{\theta}}&0\\
        0&0&\frac{\frac{\sqrt{5}}{2}-\cos{\theta}}{\sin{\theta}}
    \end{pmatrix}.
\end{equation}
Because $\rho_1$ and $\rho_2$ are both full rank and diagonal, the geodesic $\rho(t)$ is diagonal at all times. This means that it has no velocity in the coherent directions, and hence, any time-independent generator $K$, if it exists, can be made anti-Hermitian. Such a generator can thus be assumed to be of the form $K = \alpha K_\textrm{g}(0)+i\mu\id$, $\alpha \in \mathbb{R}_+, \mu\in\mathbb{R}$. 
We further show in Appendix \ref{time-independent generators} that if the generator of the shortest path is time-independent and anti-Hermitian, then it necessarily has only two distinct eigenvalues. Since this is not the case for such $K$, we must conclude that a time-independent generator that generates the shortest path does not always exist.

\subsection{Time-independent geodesic generators on \texorpdfstring{$\Q_n$}{\265} \label{time-independent generators}}
Consider a full-rank state evolution generated by a non-Hermitian time-independent operator $K = H - i \Gamma$. One possible lift of this dynamics to the purification space is described by the Schrödinger equation $\dot{W}(t) = -iHW(t) -(\Gamma -\langle \Gamma\rangle)W(t)$. Due to the fact that the images of geodesics are great arcs, we may say that if the image of $W(t)$ is an image of a geodesic, then it lies in the real span formed by the initial state $W(0)$ and the initial tangent $\dot{W}(0)$. Consequently, by Taylor expanding the evolution over some open interval and using the fact that $W$ is invertible because of the full-rank condition, we get $K^n\in\linspan\{\mathds{1},K\}$ for any power $n\in\mathbb{N}$. In this case, we can always subtract a scaling of the identity from $K$ to ensure that $K^2 =k\mathds{1}$, where $k$ is a real number\footnote{Indeed, since $K^2 = a\mathds{1} + ibK$ holds for some pair of real numbers $a,b$, we can look at $K' = K - c\mathds{1}$ for which  $K'^2 = K^2 -2cK + c^2\mathds{1} = (a+c^2)\mathds{1} + (ib-2c)K$. Taking $c = ib/2$ and $k = a - b^2/4$, we get $K'^2 = k\mathds{1}$.}. So up to rescaling, we get three cases: $K^2 = \pm\mathds{1}$ and $K^2 = 0$; thus the generator is either diagonalizable with eigenvalues $\{1,-1\}$, $\{i,-i\}$, or nilpotent of degree two.

For the geodesic to be a shortest one between the fibers, it must be horizontal. This is the case iff $W^\dagger(K^\dagger + K)W = 0$, which for invertible $W$ is equivalent to $K^\dagger + K = 0$. We hence have that the generator of the shortest geodesic on $\Q_n$ has to have its spectrum contained in $\lbrace i,-i\rbrace$, up to a global shift and rescaling.

\subsection{Qubit geodesics contained on an ellipse}
\label{ellipse}

Let $K = i\sigma_z$ and consider a general qubit state which, without loss of generality, is assumed to be $\rho = \frac{1}{2}(\mathds{1} + x\sigma_x + z\sigma_z)$. The (normalized) evolution is then given by
\begin{equation}
    \rho \mapsto \frac{1}{2(\cosh{2t} + z\sinh{2t})}(e^{2\sigma_z t} + x\sigma_x + ze^{2\sigma_z t}\sigma_z).
\end{equation}
The Bloch ball coordinates $(X,Z)$ are thus expressed in terms of $X = \frac{x}{\cosh{2t} + z\sinh{2t}}$ and $Z = \frac{z\cosh{2t} + \sinh{2t}}{\cosh{2t} + z\sinh{2t}}$. Now, note that
\begin{equation}
    \frac{X^2}{a} + Z^2 = 1 \iff a = \frac{X^2}{1-Z^2} = \frac{x^2}{1-z^2}.
\end{equation}
The fact that $a$ is positive and time independent proves that the state moves along an ellipse spanned by the $zx$-plane. We also see that the major axis is equal to one and thus touches two antipodal pure states, which are the eigenstates of $K$.

%\bibliographystyle{ieeetr}
%\bibliography{biblio_STA-OQS,articles}

\end{document}